\documentclass[journal]{IEEEtran}

\usepackage[utf8]{inputenc}
\usepackage[T1]{fontenc}

\usepackage[colorlinks,urlcolor=black,linkcolor=black,citecolor=black]{hyperref}
\usepackage{color,array,amsthm}
\usepackage{graphicx}

\usepackage{comment}
\usepackage{bm}
\usepackage{mathtools}

\usepackage{graphicx}
\usepackage{subcaption}

\usepackage{amsmath}
\usepackage{amsfonts}
\usepackage{amssymb}

\usepackage{cite}		

\usepackage{algorithm}
\usepackage{algpseudocode}
\makeatletter
\newcommand{\removelatexerror}{\let\@latex@error\@gobble}
\makeatother

\usepackage{siunitx}		
\usepackage{enumitem}		

\usepackage{pgfplots}
\pgfplotsset{compat=newest}
\usetikzlibrary{plotmarks}
\usetikzlibrary{arrows.meta}
\usepgfplotslibrary{patchplots}
\usepackage{grffile}
\pgfplotsset{plot coordinates/math parser=false}
\newlength\figureheight
\newlength\figurewidth
\usepgfplotslibrary{external}
\newcommand\norm[1]{\left\lVert#1\right\rVert}
\newcommand{\inner}[2]{{\langle}#1,#2{\rangle}}

\newcounter{relctr} 
\everydisplay\expandafter{\the\everydisplay\setcounter{relctr}{0}} 

\newcommand\labelrel[2]{%
  \begingroup
    \refstepcounter{relctr}%
    \stackrel{\textnormal{(\alph{relctr})}}{\mathstrut{#1}}%
    \originallabel{#2}%
  \endgroup
}
\AtBeginDocument{\let\originallabel\label} 

\newcommand\Item[1][]{%
  \ifx\relax#1\relax  \item \else \item[#1] \fi
  \abovedisplayskip=0pt\abovedisplayshortskip=0pt~\vspace*{-\baselineskip}}

\newtheorem{definition}{Definition}

\newtheorem*{observation}{Observation}

\begin{document}
%
\title{Synthesis of Near-Field Arrays based on Electromagnetic Inner Products}
%
%
%

\author{F. Lisi,~\IEEEmembership{Student Member,~IEEE,} ~A. Michel,~\IEEEmembership{Senior Member,~IEEE,}
        and~P. Nepa,~\IEEEmembership{Senior Member,~IEEE}
\thanks{F. Lisi, A. Michel and P. Nepa are with the Department of Information Engineering, University of Pisa, Pisa, Italy (e-mail: \href{mailto:francesco.lisi@phd.unipi.it}{francesco.lisi@phd.unipi.it}, \href{mailto:andrea.michel@unipi.it}{andrea.michel@unipi.it}, \href{mailto:paolo.nepa@unipi.it}{paolo.nepa@unipi.it}).}
\thanks{Manuscript received April 19, 2005; revised August 26, 2015.}}

%
%

\markboth{Journal of \LaTeX\ Class Files,~Vol.~14, No.~8, August~2015}%
{Shell \MakeLowercase{\textit{et al.}}: Bare Demo of IEEEtran.cls for IEEE Journals}
%

\maketitle

\begin{abstract}
Near-field antennas have been successfully adopted in several wireless applications. To exploit the high reconfigurability of array antennas, multiple synthesis techniques for arrays operating in the near-field region have been proposed. Building upon previous works on eigenmode expansions of the radiated fields, two synthesis methods for the excitations of near-field arrays based on the definition of an inner product on the electromagnetic fields are investigated: the "maximum norm" and "minimum error field norm" methods. The "maximum norm" method computes the array excitations that maximize either the active power flow through a target surface or the electric/magnetic energy stored in an assigned volume, depending on the adopted inner product. The performance of the maximum active power flow method is compared with the one of the simpler conjugate phase method. Furthermore, the limit solution achieved when the target surface reaches the far-field region is compared against the "maximum Beam Collection Efficiency" method. The "minimum error field norm" method allows to synthesize a given target field. As an example, the latter method is used to find the optimal excitation of a Plane Wave Generator with a spherical quiet zone. The effectiveness and performance of the discussed synthesis methods are validated through numerical simulations.
\end{abstract}


%
\IEEEpeerreviewmaketitle

\section{Introduction}

\IEEEPARstart{N}{ear}-field (NF) arrays have gained increasing popularity in the last decade thanks to their adoption in several relevant applications. These include microwave hyperthermia \cite{He2016}, noncontact microwave sensing \cite{Bogo2007,Stephan2007}, radio frequency identification (RFID) \cite{Siragusa2011, Buffi2010, Michel2018, Gu2019}, plane wave generators (PWG) \cite{Scattone2019_1,Scattone2019_2,Bucci2013} and wireless power transfer (WPT) systems \cite{Borgiotti1966, Geyi2010, Shinohara2011, Chou2021, Inserra2022, Geyi2021}. 

The most simple focusing technique for NF arrays consists in the conjugate phase (CP) method \cite{Buffi2012,Nepa2017}, which is based on the geometrical optics (GO) approximation. Array elements are excited with equal amplitude and a phase shift that compensates for the propagation phase delay between the element and the focal point. Although the method is suited for a single focal point, a sub-optimal solution for multiple focii can be obtained by computing the phase coefficients of the excitations for each focal point separately and then adding the computed excitations with properly selected amplitude weights \cite{Alvarez2012}. In \cite{Karimkashi2009} a Dolph-Chebyshev amplitude taper is combined with the CP method to control the side lobe level (SLL) of the electric field in the focal plane. Furthermore, by simulating a subarray of smaller dimension, the method can take into account the mutual coupling between adjacent elements even in an electrically large structure. In \cite{Bellizzi2018} the authors propose a method to compute the excitations that give the maximum uniform amplitude of a scalar field in a fixed grid of points located in the target focal region, while keeping the SLL below a predefined threshold in a different fixed grid of points. The problem is recasted into a convex optimization problem, but an additional set of phase terms must be used. Thus for a large number of points the method becomes computationally expensive. The method is extended to vector fields in \cite{Battaglia2020}. Several other constrained optimization techniques have been proposed to select the excitation of the NF array in \cite{Pino2019}.

A radically different approach to compute the excitations of NF arrays is proposed in \cite{Geyi2021}. Once the geometry of the array has been defined, one can apply the method of maximum power transmission efficiency (PTE) with a fictitious receiving antenna placed at the focal point. This technique is based on the computation of the excitation vector that maximizes the PTE between the transmitting array and the receiving antenna and requires the complete knowledge of the S parameter matrix of the system. By using multiple fictitious receiving antennas the method can be extended to obtain multiple focal points. Furthermore, since the method is based on the S parameter matrix, the mutual coupling between the transmitting elements is taken into account. In the same paper, the author describes an alternative method, called extended method for maximum PTE (EMMPTE), that does not require any fictitious antenna. Two different formulations of the EMMPTE are provided based on the electric energy stored in a set of volumes or the active power flow through a set of assigned surfaces. The advantage of these techniques is that the optimal solution can be computed in closed form by solving a generalized eigenvalue problem. 

Two interesting design methods for NF arrays have been proposed in \cite{Cicchetti2019-1}. Both methods are based on the expansion of the electromagnetic (EM) field in a set of modes, which correspond to a different combination of the excitations. The "Poynting-based" method guarantees the maximization of the active power flowing through a given surface, while the "Field-based" one minimizes the active power flow through a surface of the difference between the excited field and a target field. In \cite{Cicchetti2019-2} the latter method is used to synthesise both Airy and Bessel beams, providing an alternative to classical Bessel beam launchers \cite{Ettorre2018}, which have been successfully used in WPT systems \cite{Pako2021}. Both methods have a closed-form solution, but require the computation of several surface integrals.

Building upon the work in \cite{Cicchetti2019-1} and \cite{Geyi2021}, here we extend the two methods proposed by the former to any EM inner product. The two methods, named as "maximum norm" and "minimum error field norm" methods, are derived by only exploiting the three fundamental properties of an inner product. A different choice of the involved EM inner product corresponds to a different set of excitations and optimization objective. In particular, in Sec. \ref{sec: maximum active power flow through a surface} the "Poynting-based" method in \cite{Cicchetti2019-1} is derived as a special case of the "maximum norm" method. The solution provided by the latter method is compared with the CP method. The two solutions tend to each other when the target region is electrically small. When the target region is in the far-field (FF) region of the array, an approximate formulation of the method can be derived, which generalizes the method proposed in \cite{Oliveri2013}, that maximizes the beam collection efficiency (BCE), and we denote as "maximum BCE" method. In Sec. \ref{sec: PWG}, we exploit the constrained "minimum error field norm" method to find the excitations for a PWG. More specifically, the method provides the solution that minimizes the electric energy stored in the quiet zone (QZ) by the error field, i.e. the difference between the synthesised field and the target field, while keeping the ratio between the power flowing through the target region and the incident power above a given threshold. The proposed methods are validated by simulation: the algorithms are implemented in Matlab 2021b \cite{MATLAB} with the fields obtained via simulation with Altair FEKO v2019.2 \cite{FEKO2019.2}.

The paper is organised as follows. In Sec. \ref{sec: mathematical formulation} the "maximum norm" and "minimum error field norm" methods are described. In Sec. \ref{sec: maximum active power flow through a surface} the maximum active power flow method is described both in its general formulation and under the far-field approximation. In Sec. \ref{sec: PWG}, we apply the constrained "minimum error field norm" method to synthesise a PWG. Finally, some conclusions are drawn in Sec. \ref{sec: Conclusions}. All the detailed proof can be found in the appendices. 

\textit{Notation:} In this paper we use upper case letters $\bm{A}$ and lower case ones $\bm{a}$ to denote matrices and vectors, respectively. By $[\bm{A}]_n$ we mean the $n^{\text{th}}$ row vector of $\bm{A}$, while by $[\bm{A}]^n$ the $n^{\text{th}}$ column vector. $[\bm{a}]_n$ is the $n^{\text{th}}$ element of $\bm{a}$. The set $\{\bm{\hat{e}}_n\}_{n=1}^N$ denotes the canonical basis of $\mathbb{R}^N$, i.e. $[\bm{\hat{e}}_n]_m=\delta_{{nm}}$. The operators $(\cdot)^T$, $(\cdot)^H$ and $(\cdot)^*$ denote the transpose, conjugate transpose (hermitian) and conjugate operator, respectively. $\bm{I}_{N}$ denotes the $N \times N$ identity matrix, while $\bm{0}_{N \times M}$ denotes an all zeros $N \times M$ matrix. The absolute value of a scalar is represented by $|\cdot|$, while the euclidean norm of a vector by $\lVert \cdot \rVert$. The real and imaginary parts of a complex number are denoted by $\Re \{ \cdot \}$ and $\Im \{ \cdot \}$ respectively. In the following we are considering time harmonic fields, expressed as $F(\bm{r})$, where the associated real field can be computed as $f(\bm{r},t)=\Re\left\{F(\bm{r}) e^{j\omega_0t} \right\}$. Given $\bm{a},\bm{b}\in \mathbb{C}^3$, $\bm{a} \cdot \bm{b}$ corresponds to $\sum_{n=1}^3 a_n b_n$. 

\section{Mathematical formulation of the problem}
\label{sec: mathematical formulation}

\begin{figure}[t]
  \centering
  \includegraphics{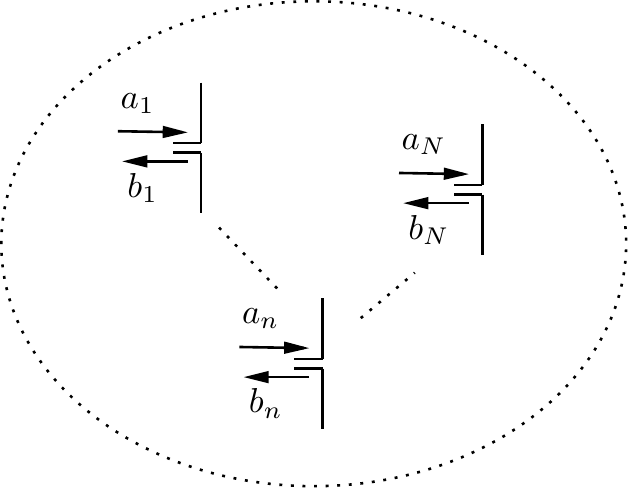}
  \caption{Schematic representation of an $N$ port array.}
  \label{fig: general array}
\end{figure}

In this work we consider an array with $N$ elements. Furthermore we suppose that each element is excited by a single mode port, thus the system can be represented as an N-port network, as depicted in Fig. \ref{fig: general array}. If we define with $\bm{a}\in \mathbb{C}^{N}$ and $\bm{b}\in \mathbb{C}^{N}$ the input and output power waves respectively, then \cite{Collin2007}
\begin{equation}
	\label{eq: S matrix}
	\bm{b}=\bm{S}\bm{a},
\end{equation}
where $\bm{S} \in \mathbb{C}^{N \times N}$ is the S parameter matrix. If the system is linear, we can express the EM field generated by the excitation vector $\bm{a}$ as
\begin{equation}
	\label{eq: EM field expansion on a basis}
	\begin{split}
		\bm{E}(\bm{r})=& \sum_{n=1}^N (\bm{\hat{x}}_n^H \bm{a})\bm{\mathcal{E}}_n(\bm{r}), \\
		\bm{H}(\bm{r})=& \sum_{n=1}^N (\bm{\hat{x}}_n^H \bm{a})\bm{\mathcal{H}}_n(\bm{r}), \\		
	\end{split}
\end{equation}
where $\big\{\bm{\hat{x}}_n \big\}_{n=1}^N$ is an orthonormal basis for the excitation space, and $(\bm{\mathcal{E}}_n,\bm{\mathcal{H}}_n)$ is the EM field generated when $\bm{a}=\bm{\hat{x}}_n$ (see Appendix \ref{ap: EM field of a linear system}). 

In the following subsections we present two methods to select the excitation vector $\bm{a}$ according to two different optimization criteria. The first method maximizes the norm of the EM field, while the second method minimizes the norm of the error field. Both methods are a generalization of the ones proposed in \cite{Cicchetti2019-1} to any EM inner product.

\subsection{Maximum norm method}
\label{sec: Maximum norm method}

Once an inner product has been selected, we can derive the excitation vector $\bm{a} \in \mathbb{C}^N$ that maximizes the induced norm for a given total incident power $\bm{a}^H\bm{B}\bm{a}$, with $\bm{B}=\bm{I}_N/2$. Otherwise one can maintain the total input power fixed as in \cite{Geyi2021}, with matrix $\bm{B}$ corresponding to \mbox{$(\bm{I}_N-\bm{S}^H\bm{S})/2$}. If the array elements are well matched to the reference impedance, and mutual coupling can be neglected, then the total input and incident power can be assumed to be equal. By using the expressions in (\ref{eq: EM field expansion on a basis}) for the electric and magnetic field, the norm squared can be expressed as
\begin{equation}
\label{eq: norm}
	\begin{split}		
		\norm{(\bm{E},\bm{H})}&^2 = \inner{(\bm{E},\bm{H})}{(\bm{E},\bm{H})}=\\
		=& \inner{\sum_{m=1}^N (\bm{\hat{x}}_m^H \bm{a})(\bm{\mathcal{E}}_m,\bm{\mathcal{H}}_m)}{\sum_{n=1}^N (\bm{\hat{x}}_n^H \bm{a})(\bm{\mathcal{E}}_n,\bm{\mathcal{H}}_n)}=\\
		=&\sum_{m=1}^N \sum_{n=1}^N ( \bm{a}^H \bm{\hat{x}}_n ) (\bm{\hat{x}}_m^H \bm{a})\inner{(\bm{\mathcal{E}}_m,\bm{\mathcal{H}}_m)}{(\bm{\mathcal{E}}_n,\bm{\mathcal{H}}_n)}=\\
		=& \bm{a}^H \bigg( \sum_{m=1}^N \sum_{n=1}^N \bm{\hat{x}}_n \bm{\hat{x}}_m^H \inner{(\bm{\mathcal{E}}_m,\bm{\mathcal{H}}_m)}{(\bm{\mathcal{E}}_n,\bm{\mathcal{H}}_n)} \bigg) \bm{a}=\\
		=&\bm{a}^H \bm{A} \bm{a},
	\end{split}
\end{equation}
where we have defined the matrix $\bm{A}$ as
\begin{equation}
	\label{def: A matrix general}
	\bm{A} \triangleq \sum_{m=1}^N \sum_{n=1}^N \bm{\hat{x}}_n \bm{\hat{x}}_m^H \inner{(\bm{\mathcal{E}}_m,\bm{\mathcal{H}}_m)}{(\bm{\mathcal{E}}_n,\bm{\mathcal{H}}_n)}=\bm{X}\bm{M}\bm{X}^H,
\end{equation}
where $\bm{X} \triangleq [\bm{\hat{x}}_1, \cdots ,\bm{\hat{x}}_N] \in \mathbb{C}^{N \times N}$ is a unitary matrix, and $\bm{M} \in \mathbb{C}^{N \times N}$ is a hermitian matrix with \mbox{$[\bm{M}]_{{nm}} \triangleq \inner{(\bm{\mathcal{E}}_m,\bm{\mathcal{H}}_m)}{(\bm{\mathcal{E}}_n,\bm{\mathcal{H}}_n)}$}. The matrix $\bm{M}$ depends on the set $\big\{(\bm{\mathcal{E}}_n,\bm{\mathcal{H}}_n) \big\}_{n=1}^N$ that in turn depends on the choice of the orthonormal basis $\big\{\bm{\hat{x}}_n \big\}_{n=1}^N$, thus $\bm{X}$ and $\bm{M}$ are set once an orthonormal basis has been chosen. By the definition of norm,  (\ref{eq: norm}) must be positive for any  $\bm{a} \in \mathbb{C}^{N}-\{\bm{0}\}$, so matrix $\bm{A}$ must be positive definite. Since $\bm{X}$ is a unitary matrix, it follows that $\bm{M}$ is positive definite too.

\begin{observation}
Let $\big\{ \bm{\hat{\varphi}}_{n},\lambda_{n} \big\}_{n=1}^N$ be the set of orthonormal eigenvectors and eigenvalues of $\bm{A}$, then its eigenvalue decomposition corresponds to
\begin{equation}
	\label{eq: EVD A}
	\bm{A}=\bm{\Phi}\bm{\Lambda}\bm{\Phi}^H.
\end{equation}
If we choose $\bm{X}=\bm{\Phi}$ and compare (\ref{def: A matrix general}) with (\ref{eq: EVD A}), it follows that $\bm{M}=\bm{\Lambda}$. Let $(\bm{\tilde{\mathcal{E}}}_{n},\bm{\tilde{\mathcal{H}}}_{n})$ be the field generated when $\bm{a}=\bm{\hat{\varphi}}_{n}$, then we can derive the following orthogonality relationship
\begin{equation}
	\label{eq: orthogonality relationship}
	\lambda_{n} \delta_{{nm}}=[\bm{\Lambda}]_{{nm}}=[\bm{M}]_{{nm}}=\inner{(\bm{\tilde{\mathcal{E}}}_{m},\bm{\tilde{\mathcal{H}}}_{m})}{(\bm{\tilde{\mathcal{E}}}_{n},\bm{\tilde{\mathcal{H}}}_{n})}.
\end{equation}
\end{observation}
Now we can solve the following constrained optimization problem
\begin{subequations}
\label{eq: constrained opt problem}
\begin{alignat}{2}
\bm{a}_{opt}=&\text{arg}\max_{\bm{a}\in \mathbb{C}^N}	&\quad	&\bm{a}^H \bm{A} \bm{a} \label{eq:optProb},\\
&\text{subject to } &	&\bm{a}^H \bm{B} \bm{a} = \bar{P}\label{eq:constraint1},
\end{alignat}
\end{subequations}
whose solution is 
\begin{equation}
	\label{eq: a opt norm}
	\bm{a}_{opt}=\sqrt{\frac{\bar{P}}{\bm{\hat\vartheta}_{max}^H \bm{B} \bm{\hat\vartheta}_{max}}} \cdot \bm{\hat\vartheta}_{max},
\end{equation}
where $\bm{\hat\vartheta}_{max}$ corresponds to the unitary eigenvector associated to the maximum eigenvalue ($\mu_{max}$) of the following generalised eigenvalue problem
\begin{equation}
	\label{eq. GEP}
	(\bm{A}-\mu\bm{B})\bm{a}=\bm{0}.
\end{equation}
A detailed proof can be found in Appendix \ref{ap: Solution of the constrained optimization problem }.


\subsection{Minimum error field norm method}
\label{sec: Minimum error field norm method}

In this section we are interested in deriving the excitation vector $\bm{a}$ that radiates the closest EM field to a target field $( \bm{\bar{E}}(\bm{r}) , \bm{\bar{H}}(\bm{r}) )$ in a specific region. Formally, we want to find the $\bm{a}$ vector that minimizes the norm of the error field, defined as
\begin{equation}
	\label{def: error field}
	\begin{split}
		&\Delta \bm{E}( \bm{r} ) \triangleq \bm{\bar{E}}( \bm{r} ) - \sum_{n=1}^N (\bm{\hat{x}}_n^H \bm{a})\bm{\mathcal{E}}_n(\bm{r}),\\
		&\Delta \bm{H}( \bm{r} ) \triangleq \bm{\bar{H}}( \bm{r} ) - \sum_{n=1}^N (\bm{\hat{x}}_n^H \bm{a})\bm{\mathcal{H}}_n(\bm{r}).
	\end{split}
\end{equation}
The solution of the optimization problem corresponds to
\begin{equation}
	\label{eq: minimum error norm a general basis}
	\bm{a}_{U}=\text{arg}\min_{\bm{a}\in \mathbb{C}^N} \norm{(\Delta \bm{E},\Delta \bm{H})}^2=\bm{A}^{-1}\bm{X}\bm{v}=\bm{X}\bm{M}^{-1}\bm{v},
\end{equation}
where we have defined $[\bm{v}]_n \triangleq \inner{(\bm{\bar{E}},\bm{\bar{H}})}{(\bm{\mathcal{E}}_n,\bm{\mathcal{H}}_n)}$, \mbox{$n=1,\dots,N$}. 
\begin{figure}[t!]
	\removelatexerror
	\begin{algorithm}[H]
	\caption{Algorithm to solve optimization problem (\ref{eq: minimum error norm constrained opt problem})}
	\label{alg: Constrained minimum error field norm method}
	\begin{algorithmic}[1]
	\State $\bm{a}_{U} \gets \bm{A}^{-1}\bm{X}\bm{v}$
	\If{$\bm{a}_{U}^H\bm{C}\bm{a}_{U}-h \leq 0$}
		\State $\bm{a}_{C} \gets \bm{a}_{U}$
	\Else
		\State Solve (\ref{eq: xi equation}) for $\xi$
		\State NormSquared $\gets +\infty$
		\For{$\xi \in \Psi_{\xi}^+$}
			\State $\bm{a} \gets (\bm{A}+\xi\bm{C})^{-1}\bm{X}\bm{v}$
			\If{$\norm{(\Delta \bm{E},\Delta \bm{H})}^2 < $ NormSquared}
				\State $\bm{a}_{C} \gets \bm{a}$
				\State NormSquared $\gets$ $\norm{(\Delta \bm{E},\Delta \bm{H})}^2$
			\EndIf
		\EndFor
	\EndIf
	\end{algorithmic}
	\end{algorithm}
\end{figure}

The main drawback of the proposed solution consists in considering the target region only, without taking into account the behaviour of the fields in the surroundings. To overcome this issue, the following constrained optimization problem can be considered
\begin{subequations}
\label{eq: minimum error norm constrained opt problem}
\begin{alignat}{2}
\bm{a}_{C}=&\text{arg}\min_{\bm{a}\in \mathbb{C}^N} &\quad &\norm{(\Delta \bm{E},\Delta \bm{H})}^2,\\
&\text{subject to } &	&\bm{a}^H \bm{C} \bm{a} -h \leq 0, \label{eq: minimum error norm constrained opt problem (constraint)}
\end{alignat}
\end{subequations}
where $h \in \mathbb{R}$ and $\bm{C} \in \mathbb{C}^{N \times N}$ is a hermitian matrix. The family of constraints expressed as (\ref{eq: minimum error norm constrained opt problem (constraint)}) includes several physically meaningful ones. One specific example will be provided in Sec. \ref{sec: PWG} where we compute the optimal excitation array for a PWG. Algorithm \ref{alg: Constrained minimum error field norm method} describes the steps to find the solution of the constrained optimization problem (\ref{eq: minimum error norm constrained opt problem}). In words, after computing the optimal solution of the unconstrained optimization problem $\bm{a}_U$, the algorithm checks if it satisfies the constraint (\ref{eq: minimum error norm constrained opt problem (constraint)}): if it is satisfied then the optimal solution corresponds to $\bm{a}_U$, otherwise the algorithm performs the following steps. After numerically solving  (\ref{eq: xi equation}), the algorithm performs a for loop over the subset of positive real solutions $\Psi_{\xi}^+$. For each value of $\xi$ the associated excitation vector $\bm{a}$ is computed as $(\bm{A}+\xi\bm{C})^{-1}\bm{X}\bm{v}$, and the associated $\norm{(\Delta \bm{E},\Delta \bm{H})}^2$ value is computed as in (\ref{eq: error field norm (a explicit) final}). The optimal excitation corresponds to the one associated to the smallest $\norm{(\Delta \bm{E},\Delta \bm{H})}^2$ value. A detailed proof of the algorithm can be found in Appendix \ref{ap: Solution of the minimum error norm optimization problem}.


\section{maximum active power flow through a surface}
\label{sec: maximum active power flow through a surface}

In this section we apply the "maximum norm" method described in Sec. \ref{sec: Maximum norm method} with the inner product defined in Sec. \ref{sec: maximum active power flow: General case}, whose induced norm corresponds to the active power flow through a surface. As a consequence the "maximum norm" method provides the excitation array that maximizes the power flow through a given surface and corresponds to the "Poynting-based" method in \cite{Cicchetti2019-1}. Under the assumption that the surface belongs to the far-field region of the array, in Sec. \ref{sec: maximum active power flow: Far-Field case} we derive an explicit formulation of matrix $\bm{A}$ as a function of the active pattern of each element.  Furthermore by considering ideal isotropic radiators, the above method converges to the  "maximum BCE" method in \cite{Oliveri2013}. In Sec. \ref{sec: maximum active power flow: general case simulation results} and Sec. \ref{sec: maximum active power flow: Far-Field case simulation results}, the results are validated via numerical simulation.


\subsection{maximum active power flow: General case}
\label{sec: maximum active power flow: General case}

\begin{figure}[t]
  \centering
  \includegraphics{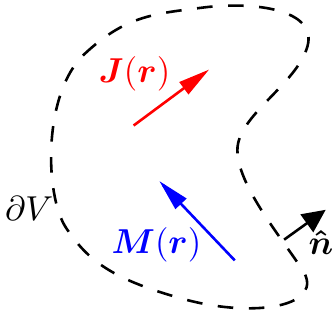}
  \caption{Closed surface $\partial V$ with oriented unit vector $\bm{\hat{n}}$.}
  \label{fig: closed surface}
\end{figure}

First we start by defining the following set.

\begin{definition}[O set]
\label{def: O set}
Given a surface $S$ and its unit normal vector $\bm{\hat{n}}$, we define the set 
\begin{equation}
	\mathcal{O}(S) \triangleq \bigg\{ (\bm{E},\bm{H}) \neq (\bm{0},\bm{0}): \frac{1}{2} \iint_{S} \Re \big\{ \bm{E} \times \bm{H}^* \big\} \cdot \bm{\hat{n}}d\Sigma > 0 \bigg\},
\end{equation}
containing all the EM fields that satisfy the Maxwell equations in the neighbourhood of $S$, that have a positive active power flow through $S$ in the direction given by $\bm{\hat{n}}$. 
\end{definition}

As an example, we consider a volume $V$ and its boundary $\partial V$ with normal unit vector in the outward direction, as depicted in Fig. \ref{fig: closed surface}. In this case the set $\mathcal{O}(\partial V)$ contains all the EM fields that are generated by sources inside the volume $V$. Now we can define the following inner product.

\begin{definition}[EM inner product]
\label{def: EM inner product}
Given two sets of solution of Maxwell equations $(\bm{E}_1,\bm{H}_1),(\bm{E}_2,\bm{H}_2) \in \mathcal{O}(S)$, the following operator
\begin{equation}
	\label{eq: EM inner product definition}
	\begin{split}
		\inner{(\bm{E}_1,\bm{H}_1)}{&(\bm{E}_2,\bm{H}_2)} \triangleq \\
		&\triangleq \frac{1}{4} \iint_{S} (\bm{E}_1 \times \bm{H}_2^* + \bm{E}_2^* \times \bm{H}_1) \cdot \bm{\hat{n}}d\Sigma,
	\end{split}
\end{equation}
satisfies the properties of an inner product.
\end{definition}

\noindent \textit{Proof.} see Appendix \ref{ap: Proof of the properties of the EM inner product}.\\

This is one possible choice of an inner product on EM fields. The norm induced from this inner product corresponds to
\begin{equation}
	\label{eq: norm as power flow}
	\begin{split}
		\norm{(\bm{E},\bm{H})}^2 =& \inner{(\bm{E},\bm{H})}{(\bm{E},\bm{H})}=\\
		=&\frac{1}{4} \iint_{S} (\bm{E} \times \bm{H}^* + \bm{E}^* \times \bm{H}) \cdot \bm{\hat{n}}d\Sigma=\\
		=&\frac{1}{2} \iint_{S} \Re \left\{\bm{E} \times \bm{H}^* \right\} \cdot \bm{\hat{n}}d\Sigma =\bm{a}^H \bm{A} \bm{a},
	\end{split}
\end{equation}
and 
\begin{equation}
	\label{eq: M matrix (power flow)}
	\begin{split}
		[\bm{M}]_{{nm}}=\frac{1}{4} \iint_{S} (\bm{\mathcal{E}}_m \times \bm{\mathcal{H}}_n^* + \bm{\mathcal{E}}_n^* \times \bm{\mathcal{H}}_m) \cdot \bm{\hat{n}}d\Sigma
	\end{split}
\end{equation}
where $\bm{A}=\bm{X}\bm{M}\bm{X}^H$.
Thus the optimal excitation vector in (\ref{eq: a opt norm}) corresponds to the one that maximizes the radiated power through the surface $S$, when the total incident power is $\bar{P}$. This is exactly the same result as the one derived in \cite{Cicchetti2019-1}, where the authors refer to it as "Poynting-based" technique. Furthermore the orthogonality relationship in (\ref{eq: orthogonality relationship}) becomes
\begin{equation}
	\label{eq: ort relationship (power flow)}
	\frac{1}{4} \iint_{S} (\bm{\tilde{\mathcal{E}}}_m \times \bm{\tilde{\mathcal{H}}}_n^* + \bm{\tilde{\mathcal{E}}}_n^* \times \bm{\tilde{\mathcal{H}}}_m) \cdot \bm{\hat{n}}d\Sigma=\lambda_{n} \delta_{{nm}},
\end{equation}
that is the same as Eq. (18) in \cite{Cicchetti2019-1}, apart from a factor $2$ that is accounted by $\lambda_n$ in this paper.

\begin{figure}[t!]
  \centering
  \includegraphics{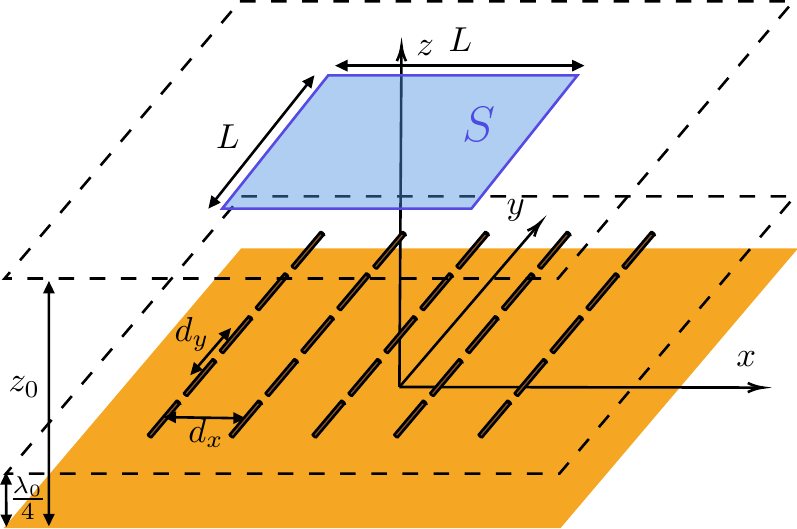}
  \caption{$y$-oriented half-wave dipole array placed above an infinite ground plane. The square surface $S$ corresponds to the region where the radiated power must be maximized.}
  \label{fig: Schematic_Array_5x5_NoCoupling_FixedZ}
\end{figure}

If, instead, we consider the following inner product
\begin{equation}
	\label{def: Geyi power flow based inner product}
	\begin{split}
		\inner{(\bm{E}_1,\bm{H}_1)}{(\bm{E}_2,\bm{H}_2)} & \triangleq \\
		\triangleq \frac{1}{4} \iint_{\bigcup\limits_{p=1}^P S_p} & W(\bm{r})\bigl[\bm{E}_1(\bm{r}) \times \bm{H}_2^*(\bm{r}) +\\
		& + \bm{E}_2^*(\bm{r}) \times \bm{H}_1(\bm{r}) \bigr] \cdot \bm{\hat{n}}d\Sigma=\\
		= \frac{1}{4} \sum_{p=1}^P \iint_{S_p} & W(\bm{r})\bigl[\bm{E}_1(\bm{r}) \times \bm{H}_2^*(\bm{r}) +\\
		& + \bm{E}_2^*(\bm{r}) \times \bm{H}_1(\bm{r}) \bigr] \cdot \bm{\hat{n}}d\Sigma,
	\end{split}
\end{equation}
where $\{ S_p \}_{p=1}^P$ is a set of separate target surfaces, we obtain the EMMPTE proposed in Sec. III.B of \cite{Geyi2021}. $W(\bm{r})\in \mathbb{R}$ is an optional weighting function and must be chosen in such a way that the inner product in (\ref{def: Geyi power flow based inner product}) is positive-definite.


\subsection{maximum active power flow: general case simulation results}
\label{sec: maximum active power flow: general case simulation results}

\begin{figure}[t!]
  \centering
  \includegraphics{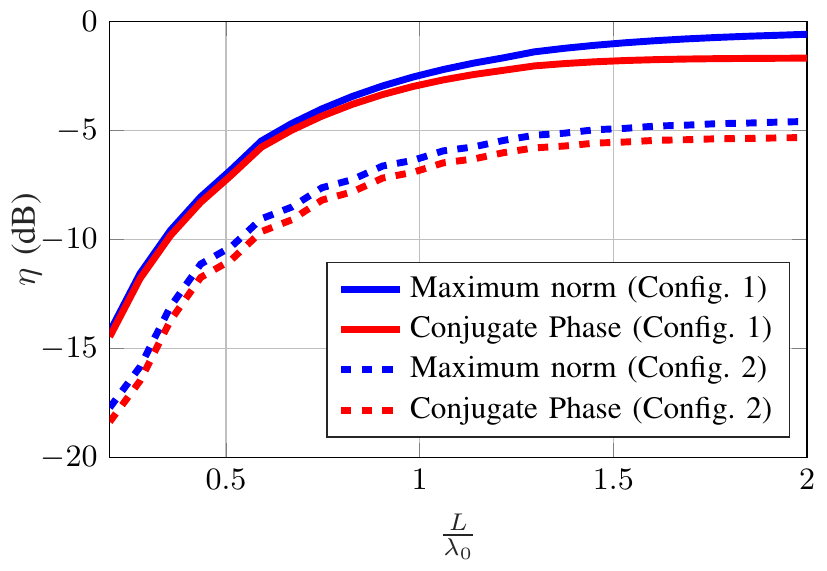}
  \caption{Plot of the efficiency $\eta=P_{S}/P_{inc}$ as a function of the square surface $S$ side length $L$. The solid curves correspond to Configuration 1 (negligible coupling effects), the dashed curves to Configuration 2 (stronger coupling effects).}
  \label{fig: Array_FixedZ}
\end{figure}
\begin{table}[t!]
\caption{Parameters of the two simulated configurations.}
\label{table:Parameters configuration (coupling and no coupling)}
\centering
\def\arraystretch{1.2}
	\begin{tabular}{*{5}{|c}|}

 	\hline
 	Configuration & $N_x$ & $d_x$ & $N_y$ & $d_y$ \\ 
 	\hline
 	1 & 4 & $0.6\lambda_0$ & 4 & $0.6\lambda_0$  \\ 
 	\hline
 	2 & 19 & $0.1\lambda_0$ & 4 & $0.6\lambda_0$  \\ 
	\hline
	\end{tabular}
\end{table}

In this section we present the results obtained by simulating the configuration shown in Fig. \ref{fig: Schematic_Array_5x5_NoCoupling_FixedZ}. The array consists of \mbox{$N_x \times N_y$} half-wave dipoles placed $\lambda_0/4$ above an infinite ground plane resonating at $1$ \si{\giga\hertz}. The dipoles are $142.1$ \si{\milli\metre} long with radius equal to $30$ \si{\micro\metre} and excitation gap $0.3$ \si{\milli\metre}. $d_x$ and $d_y$ represent the inter-element spacing along the $x$ and $y$ direction, respectively. The surface $S$ is a square of length $L$ in the plane $z=z_0=2\lambda_0$. Fig. \ref{fig: Array_FixedZ} shows the efficiency $\eta=P_{S}/P_{inc}$, where $P_{S}$ is the active power flow through the surface $S$ in the $z$ direction and $P_{inc}$ is the total incident power. The "maximum norm" method is compared to the CP method \cite{Nepa2017} for two different array configurations. Table \ref{table:Parameters configuration (coupling and no coupling)} lists the parameters of the two configurations: both arrays occupy the same area, but in Configuration 2 the coupling between each element and the adjacent ones in the $x$ direction is above $-10$ dB, while in Configuration 1 the coupling effects can be neglected. When the sides of the surface  $S$ are small compared to a wavelength the performance of the CP method approaches the one of the "maximum norm" method. When $L=0.2\lambda_0$, the difference between the efficiency of the two methods is $0.22$ dB and $0.65$ dB for Configuration 1 and 2, respectively. The larger difference for Configuration 2 is due to the stronger effects of the coupling, which is neglected by the CP method. The "maximum norm" method outperforms the CP method in both configurations. Since the CP method focuses the field at $(0,0,z_0)$, while the "maximum norm" method maximizes the power flow through $S$, the difference in performance between the two synthesis techniques increases for larger arrays. Due to reflection losses induced by a stronger EM coupling, Configuration 2 loses around $4$ dB compared to Configuration 1.


\subsection{maximum active power flow: Far-Field case}
\label{sec: maximum active power flow: Far-Field case}

\begin{figure}[t!]
  \centering
  \includegraphics{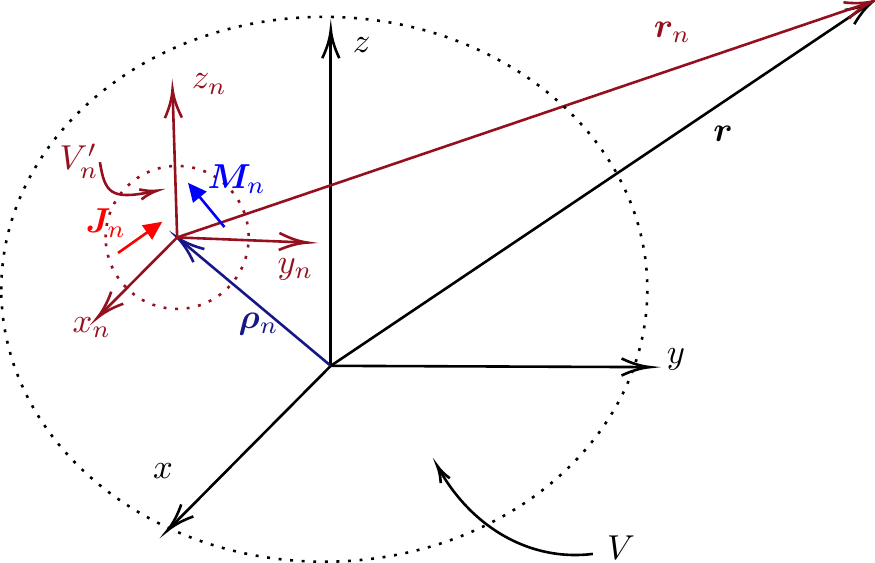}
  \caption{Relative coordinate system centred at the position of the $n^{\text{th}}$ array element $\bm{\rho}_n$. $V$ is a volume that bounds all the sources of the array in the global coordinate system. $V'_n$ is a volume that bounds the sources of the $n^{\text{th}}$ array element in its relative coordinate system. }
  \label{fig: GCS and LCS}
\end{figure}

In this section we suppose that the surface $S$ is located in the far-field region of the array. Furthermore, we assume that the mutual coupling between the array elements can be neglected. Finally, we consider the canonical basis as the orthonormal basis, so $\bm{X}=\bm{I}_N$. In the following, we use the symbol $\check{(\cdot)}$ to denote a quantity evaluated using the canonical basis $\bm{a}=\bm{\hat{e}}_n$. By referring to Fig. \ref{fig: GCS and LCS}, $\bm{J}_n(\bm{a},\bm{r}_n)$ and $\bm{M}_n(\bm{a},\bm{r}_n)$ denote the electric and magnetic current densities within volume $V'_n$ at the $n^{\text{th}}$ array element in its local coordinate system when the excitation vector is equal to $\bm{a}$. Thus $\bm{\check{J}}_n(\bm{r}_n) \triangleq \bm{J}_n(\bm{\hat{e}}_n,\bm{r}_n)$ and $\bm{\check{M}}_n(\bm{r}_n) \triangleq \bm{M}_n(\bm{\hat{e}}_n,\bm{r}_n)$. In the general case when the EM coupling effects cannot be neglected, the electric and magnetic current of element $n$ depend on the whole excitation vector $\bm{a}$. The electric field radiated by the $n^{\text{th}}$ array element at $\bm{r}$, which lies in the far-field region of the whole array, in the global coordinate system reference can be expressed as \cite{Balanis2015}
\begin{equation}
	\label{eq: EF element n far field (a)}
	\begin{split}
		\bm{\check{\mathcal{E}}}_n(\bm{r})=&j\omega [-\bm{\hat{r}} \times \bm{\check{A}}_n(\bm{r}) \times \bm{\hat{r}} +\zeta \bm{\hat{r}} \times \bm{\check{F}}_n(\bm{r})]= \\
		=&j\zeta \frac{k}{4\pi} \frac{e^{-jkr}}{r} \iiint_{V'_n} \biggl[ -\bm{\hat{r}} \times \bm{\check{J}}_n(\bm{r}'_n) \times \bm{\hat{r}} + \\
		& \quad\quad +  \frac{1}{\zeta} \bm{\hat{r}} \times \bm{\check{M}}_n(\bm{r}'_n) \biggr] e^{jk \bm{\hat{r}} \cdot (\bm{r}'_n+\bm{\rho}_n)} d\tau'= \\
		=& j \frac{\zeta}{2\lambda} \frac{e^{-jkr}}{r} \bm{\check{g}}_n(\bm{\hat{r}}) e^{jk\bm{\hat{r}} \cdot \bm{\rho}_n},
	\end{split}
\end{equation}
where we have defined
\begin{equation}
	\label{def: g array}
	\bm{\check{g}}_n(\bm{\hat{r}}) \triangleq \iiint_{V'_n} \bm{\hat{r}} \times \biggl[ -\bm{\check{J}}_n(\bm{r}'_n) \times \bm{\hat{r}} +  \frac{1}{\zeta} \bm{\check{M}}_n(\bm{r}'_n) \biggr] e^{jk \bm{\hat{r}} \cdot \bm{r}'_n} d\tau',
\end{equation}
and $\bm{\check{\mathcal{H}}}_n(\bm{r})=[\bm{\hat{r}} \times \bm{\check{\mathcal{E}}}_n(\bm{r})]/\zeta$. $\bm{\check{A}}_n(\bm{r})$ and $\bm{\check{F}}_n(\bm{r})$ are the electric and magnetic vector potentials, $k$, $\lambda$ and $\zeta$ are the wavenumber, wavelength and characteristic impedance of free space, \mbox{$r \triangleq \norm{\bm{r}}$} and \mbox{$\bm{\hat{r}}\triangleq \bm{r}/r$}. By substituting the expressions of $\bm{\check{\mathcal{E}}}_n$ and $\bm{\check{\mathcal{H}}}_n$ into (\ref{eq: M matrix (power flow)}) and $\bm{X}=\bm{I}_N$ in (\ref{def: A matrix general}), we obtain
\begin{equation}
	\label{eq: A matrix (far field)}
	\begin{split}
	[\bm{A}]_{{nm}}=&[\bm{M}]_{{nm}}=\frac{1}{4} \iint_{S} (\bm{\check{\mathcal{E}}}_m \times \bm{\check{\mathcal{H}}}_n^* + \bm{\check{\mathcal{E}}}_n^* \times \bm{\check{\mathcal{H}}}_m) \cdot \bm{\hat{n}}d\Sigma=\\
	=& \frac{1}{2\zeta} \iint_{S} (\bm{\check{\mathcal{E}}}_m \cdot \bm{\check{\mathcal{E}}}^*_n) (\bm{\hat{r}} \cdot \bm{\hat{n}}) d\Sigma=\\
	=& \frac{\zeta}{8\lambda^2} \iint_{S} \bm{\check{g}}_m(\bm{\hat{r}}) \cdot \bm{\check{g}}^*_n(\bm{\hat{r}}) e^{jk \bm{\hat{r}} \cdot (\bm{\rho}_m - \bm{\rho}_n)} \frac{\bm{\hat{r}} \cdot \bm{\hat{n}}}{r^2} d\Sigma,
	\end{split}
\end{equation}
By considering a planar array in the $x$-$y$ plane, Eq. (\ref{eq: A matrix (far field)}) can be expressed in spherical coordinates as
\begin{equation}
	\label{eq: A matrix (far field, planar array)}
	\begin{split}
		[\bm{A}]_{{nm}}=&\frac{\zeta}{8\lambda^2} \iint_{\Omega} e^{jk \sin(\theta) [ (x_m - x_n)\cos(\phi)+(y_m - y_n)\sin(\phi)]} \\
		& \qquad\qquad\qquad \bm{\check{g}}_m(\theta,\phi) \cdot \bm{\check{g}}^*_n(\theta,\phi) \sin(\theta)d\theta d\phi,
	\end{split}
\end{equation} 
where $\bm{\rho}_n=x_n\bm{\hat{x}}+y_n\bm{\hat{y}}$, $n=1,\dots,N$, and $\Omega$ is the subtended solid angle.

\begin{figure}[t!]
  \centering
  \includegraphics{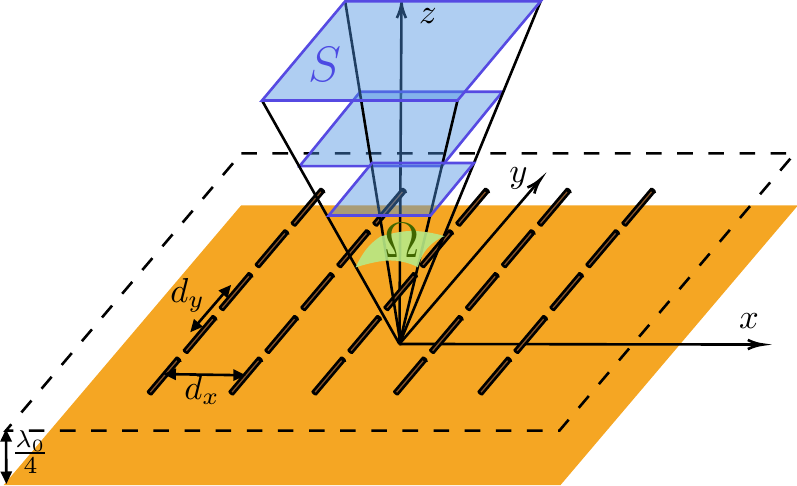}
  \caption{$y$-oriented half-wave dipole array placed above an infinite ground plane. All the surfaces $S$ subtend the same solid angle $\Omega$.}
  \label{fig: Schematic_Array_5x5_NoCoupling_FixedSolidAngle}
\end{figure}
Finally, to verify that the far-field approximation of the "maximum norm" method coincides with the "maximum BCE" method in \cite{Oliveri2013}, we assume that all the elements are ideal isotropic radiators and there are no losses. Under the above assumptions the same results as in \cite{Oliveri2013} are obtained:
\begin{equation}
	\label{eq: A and B (FF, isotropical)}
	\begin{split}
		[\bm{A}]_{{nm}}=&\frac{\zeta}{8\lambda^2}\norm{\bm{\check{g}}}^2 \iint_{\Omega} e^{jk \sin(\theta) [ \Delta x_{nm} \cos(\phi)+\Delta y_{nm}\sin(\phi)]} \\
		& \qquad\qquad\qquad\qquad  \sin(\theta)d\theta d\phi,\\
		[\bm{B}]_{{nm}}=&\frac{\zeta}{8\lambda^2}\norm{\bm{\check{g}}}^2 \int_{0}^{2\pi}\int_{0}^{\pi} e^{jk \sin(\theta) [ \Delta x_{nm} \cos(\phi)+\Delta y_{nm}\sin(\phi)]} \\
		& \qquad\qquad\qquad\qquad  \sin(\theta)d\theta d\phi=\\
		=& \frac{\zeta}{8\lambda^2}\norm{\bm{\check{g}}}^2 4\pi \frac{\sin(k\sqrt{\Delta x_{nm}^2+\Delta y_{nm}^2})}{k\sqrt{\Delta x_{nm}^2+\Delta y_{nm}^2}},
	\end{split}
\end{equation} 
where we have defined $\Delta x_{nm} \triangleq x_m-x_n$ and $\Delta y_{nm} \triangleq y_m-y_n$. By computing the solution in (\ref{eq: a opt norm}) with the expressions in (\ref{eq: A and B (FF, isotropical)}) for $\bm{A}$ and $\bm{B}$, this corresponds to the "maximum BCE" method. Thus the "maximum BCE" method is a special case of the "maximum norm" method. Since the multiplicative term before the integrand in (\ref{eq: a opt norm}) is the same for $\bm{A}$ and $\bm{B}$, the optimal solution does not depend on its value. The "maximum BCE" method does not require any information about the radiating properties of the elements of the array and depends solely on the selected surface $S$. 


\subsection{maximum active power flow: Far-Field case simulation results}
\label{sec: maximum active power flow: Far-Field case simulation results}

\begin{figure}[t!]
  \centering
  \includegraphics{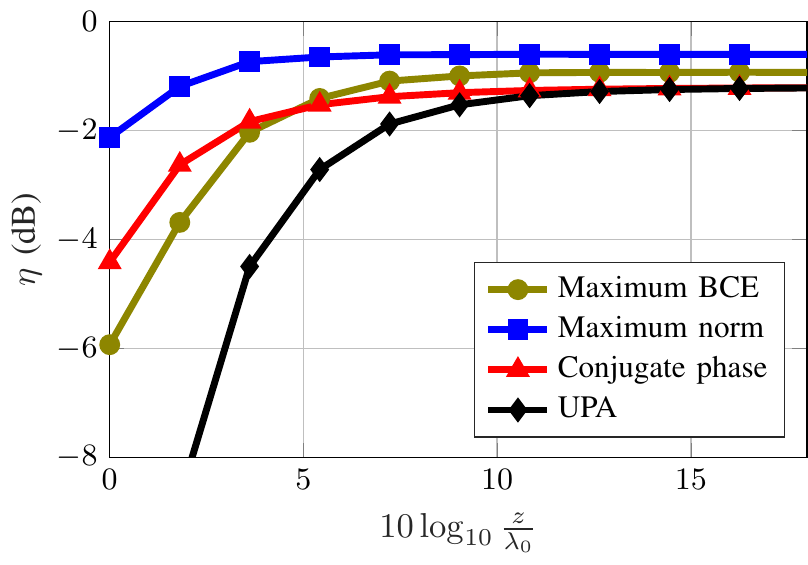}
  	\caption{Plot of the efficiency $\eta=P_{S}/P_{inc}$ as a function of the distance $z$ between the surface $S$ and the infinite ground plane below the array of dipoles. The solid angle $\Omega$ subtended by the surface $S$ is the same for every $z$ (see Fig. \ref{fig: Schematic_Array_5x5_NoCoupling_FixedSolidAngle}).}
   	\label{fig: Array_5x5_NoCoupling_FixedSolidAngle}
\end{figure}

In this section we validate the results found in Sec. \ref{sec: maximum active power flow: Far-Field case} by simulating a square array of $5 \times 5$ half-wave dipoles as the ones in Sec. \ref{sec: maximum active power flow: general case simulation results}. The inter-element spacing is equal to $0.6\lambda_0$ in both directions. We consider the efficiency of various methods when the surface $S$ moves along the $z$ direction while keeping its subtended solid angle $\Omega$ constant, as shown in Fig. \ref{fig: Array_5x5_NoCoupling_FixedSolidAngle}. The length of the side of the square surface $S$ at $z$ can be computed as $L=2z\tan(\theta_0)$. The $\theta_0$ value has been chosen as the first null angle of the array factor of an uniform planar array (UPA), i.e. $\theta_0=\text{asin}(\lambda_0/(N_xd_x))$. The Fraunhofer distance is approximately $D_{F}=2((N_xd_x)^2+(N_yd_y)^2)/\lambda_0=36\lambda_0$. Fig. \ref{fig: Array_5x5_NoCoupling_FixedSolidAngle} shows the performance in terms of efficiency of the following four algorithms: the general "maximum norm", the "maximum BCE", the UPA and the CP method with an assigned focal point at the center of the surface. When the surface $S$ is closer to the array plane the general "maximum norm" method and the CP method have better performance compared to the "maximum PTE" and the UPA, since the latter methods are based on the far-field approximation of the fields. However when the distance grows, the "maximum BCE" method performance surpasses the CP one. Since the "maximum BCE" method does not take into account the element factor, as described in Sec. \ref{sec: maximum active power flow: Far-Field case}, there is still a gap between its performance and the one of the "maximum norm" method, when the surface $S$ is in the far-field region. The performance of the UPA tend to the one of the CP method in the far-field, as expected. As a final remark, the performance gap between the "maximum norm" ("maximum BCE") and the CP (UPA) method in the near-field (far-field) increases when considering a wider solid angle $\Omega$. This is confirmed by the results in Sec. \ref{sec: maximum active power flow: general case simulation results} for the near-field region.


\section{Minimum error field norm method: Plane Wave Generator excitations synthesis}
\label{sec: PWG}

\begin{figure}[t!]
  \centering
  \includegraphics{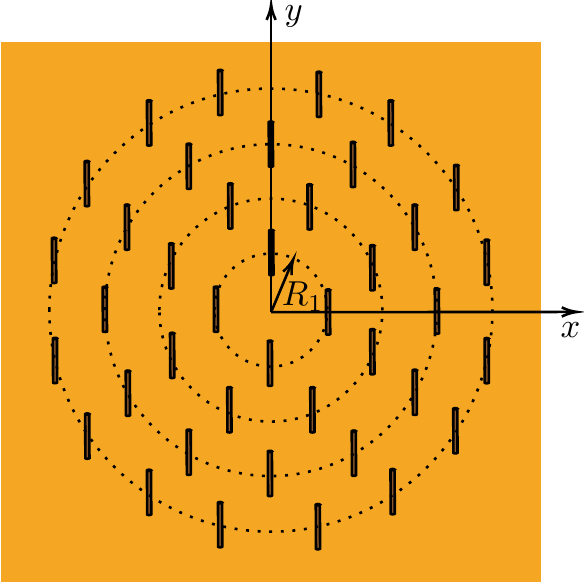}
  \caption{Top view of the simulated PWG. The PWG consists of 40 half-wave dipoles oriented in the y direction and disposed in four concentrical rings.}
  \label{fig: PWG schematic}
\end{figure}

In this section we exploit the "minimum error field norm" method to select the optimal excitations for the synthesis of a PWG. Since obtaining a linearly polarized plane wave in a spherical region of radius $R$ and centered at $(0,0,z_0)$ is our objective, we exploit the following volumetric inner product
\begin{equation}
\label{eq: PWG inner product}
	\inner{(\bm{E}_1,\bm{H}_1)}{(\bm{E}_2,\bm{H}_2)} \triangleq \iiint_{V} \frac{1}{4}\bm{E}_1(\bm{r}) \cdot \bm{E}_2^*(\bm{r})d\tau,
\end{equation}
where $V$ is the target spherical region or QZ.
The target field corresponds to a linearly polarized plane wave travelling in the direction normal to the planar array, that we assume to be the $z$ direction, i.e. \mbox{$\bm{\bar{E}}=E_0e^{-jk(z-z_0)} \bm{\hat{y}}$} and \mbox{$\bm{\bar{H}}=-E_0e^{-jk(z-z_0)} \bm{\hat{x}}/\zeta$}. As a consequence the elements of $\bm{M}$ and $\bm{v}$, as defined in Sec. \ref{sec: Maximum norm method} and \ref{sec: Minimum error field norm method} respectively, correspond to 
\begin{equation}
	\label{eq: PWG M and v}
	\begin{split}
		[\bm{M}]_{nm}=&\inner{(\bm{\mathcal{E}}_{m},\bm{\mathcal{H}}_{m})}{(\bm{\mathcal{E}}_{n},\bm{\mathcal{H}}_{n})}=\iiint_{V} \frac{\bm{\mathcal{E}}_m \cdot \bm{\mathcal{E}}_n^*}{4}d\tau, \\
		[\bm{v}]_n=&\inner{(\bm{\bar{E}},\bm{\bar{H}})}{(\bm{\mathcal{E}}_n,\bm{\mathcal{H}}_n)}=\iiint_{V} \frac{\bm{\bar{E}} \cdot \bm{\mathcal{E}}_n^*}{4}d\tau,
	\end{split}
\end{equation}
and $\bm{A}=\bm{X}\bm{M}\bm{X}^H$. These are all the elements needed to compute the excitation vector with the "minimum error field norm" method without constraints, as described in Sec. \ref{sec: Minimum error field norm method}. The main drawback of this technique is that the optimization problem depends only on the properties of the solution in the QZ, without considering the fields generated in the surroundings. To overcome this problem, we consider the solution that satisfies the following constraint: the active power flow through the circular surface $S$, which is obtained by cutting the spherical QZ with a plane parallel to the array plane passing through the QZ center, must be greater than the incident power divided by a factor $\alpha > 1$. As in Sec. \ref{sec: maximum active power flow: General case}, the power flowing through the surface $S$ can be expressed as $P_S=\bm{a}^H\bm{C}_1\bm{a}$ with $\bm{C}_1=\bm{X}\bm{M}_1\bm{X}^H$ and 
\begin{equation}
	\label{eq: PWG M1}
		[\bm{M}_1]_{nm}=\frac{1}{4} \iint_{S} (\bm{\mathcal{E}}_m \times \bm{\mathcal{H}}_n^* + \bm{\mathcal{E}}_n^* \times \bm{\mathcal{H}}_m) \cdot \bm{\hat{n}}d\Sigma,
\end{equation}
where $\bm{C}_1$ and $\bm{M}_1$ correspond to $\bm{A}$ and $\bm{M}$ in Sec. \ref{sec: maximum active power flow: General case}. The incident power corresponds to $P_{inc}=\bm{a}^H\bm{B}\bm{a}$ with $\bm{B}=\bm{I}_N/2$, thus we can express the constraint as $P_S \geq  P_{inc}/ \alpha $ or equivalently
\begin{equation}
	\label{eq: PWG constraint}
	  P_{inc}-\alpha P_S= \bm{a}^H(\bm{B}-\alpha \bm{C}_1)\bm{a}=\bm{a}^H \bm{C}\bm{a} \leq 0
\end{equation}
\begin{figure}[t!]
  \centering
  \includegraphics{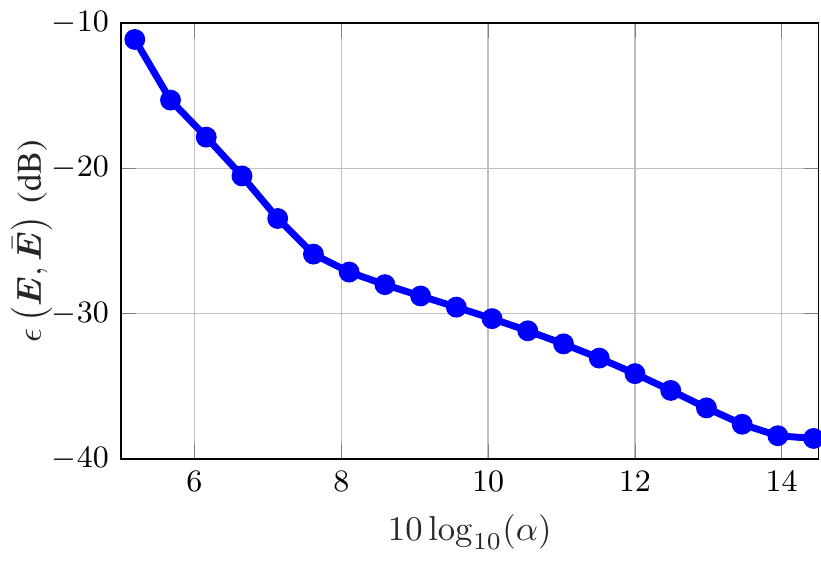}
  \caption{Relative error, as defined in (\ref{eq: PWG relative error}), of the PWG field as a function of the parameter $\alpha$.}
  \label{fig: PWG norm}
\end{figure}

where we have defined the matrix $\bm{C}\triangleq\bm{B}-\alpha \bm{C}_1$. As a consequence the constraint belongs to the family defined by (\ref{eq: minimum error norm constrained opt problem (constraint)}) with $h=0$. Since quantities related to power and energy can be expressed as bilinear forms of $\bm{a}$ in a linear system, it is now clear why the considered family of constraints contains several physically relevant ones. It is worth mentioning that, given the array geometry, the $\alpha$ value cannot be chosen arbitrarily. A lower bound on $\alpha$ can be found from the following inequality chain
\begin{equation}
	\label{eq: PWG minimum alpha}
	\alpha_{min} \leq \frac{\bm{a}^H\bm{B}\bm{a}}{\bm{a}^H\bm{C}_1\bm{a}} \leq \alpha
\end{equation}
where the last inequality follows from (\ref{eq: PWG constraint}), and the first one from the properties of Rayleigh quotients. Since the expression in the middle corresponds to a generalized Rayleigh quotient, then $\alpha_{min}$ is the smallest eigenvalue of the generalized eigenvalue problem $\bm{B}\bm{a}=\mu\bm{C}_1 \bm{a}$. The physical meaning of the lower bound on $\alpha$ can be explained as follows: once the geometry of the array and the surface $S$ are assigned, there is a maximum amount of active power that can flow through the surface for a given input power, as discussed in Sec. \ref{sec: maximum active power flow through a surface}. Likewise, the upper bound on $\alpha$ corresponds to $\alpha_{max}=(\bm{a}_U^H\bm{B}\bm{a}_U)/(\bm{a}_U^H\bm{C}_1\bm{a}_U)$. In fact if we select $\alpha > \alpha_{max}$ then 
\begin{equation}
\label{eq: PWG alpha max}
	\bm{a}_U^H\bm{C}\bm{a}_U=\bm{a}_U^H(\bm{B}-\alpha \bm{C}_1)\bm{a}_U < \bm{a}_U^H(\bm{B}-\alpha_{max} \bm{C}_1)\bm{a}_U=0.
\end{equation}
\begin{figure*}[t!]
  	\centering
    \begin{subfigure}[b]{0.475\linewidth}   
      	\centering 
        \includegraphics{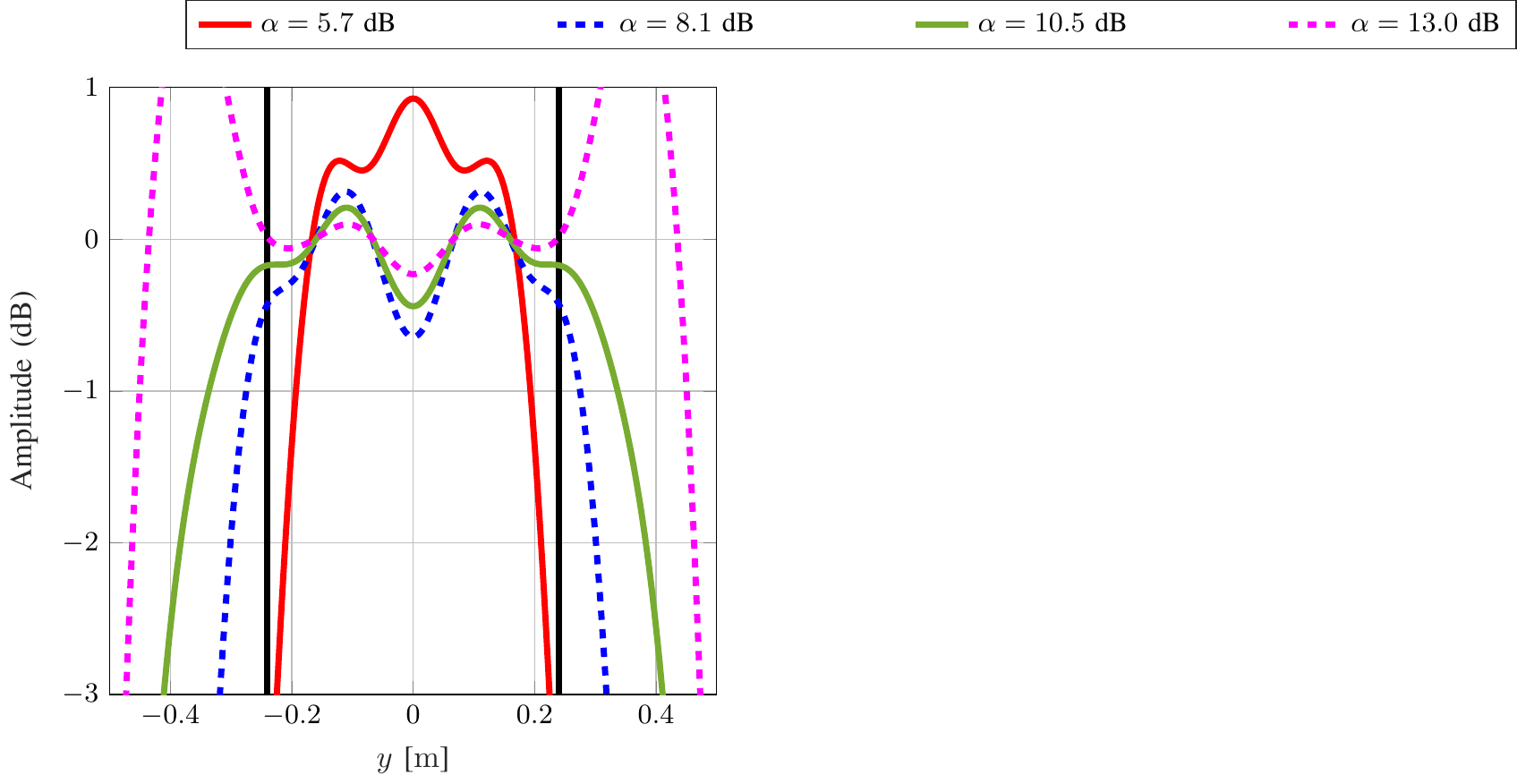}
        \caption[]%
        {{\small Amplitude}}    
        \label{fig: PWG 1D amp}
   	\end{subfigure}
    \hfill
    \begin{subfigure}[b]{0.475\linewidth}   
      	\centering 
        \includegraphics{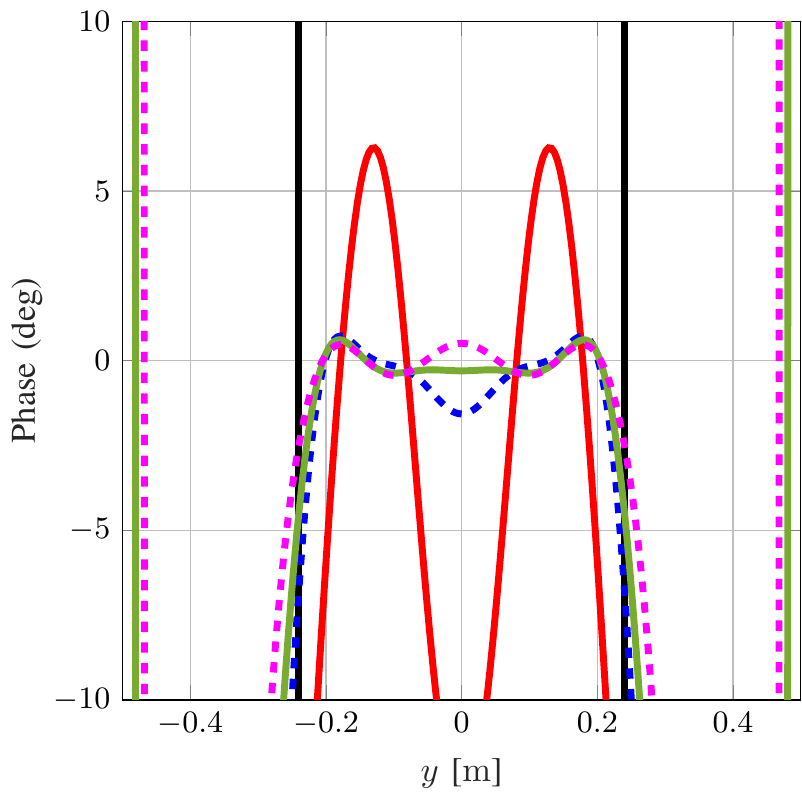}
        \caption[]%
        {{\small Phase}}    
        \label{fig: PWG 1D phase}
  	\end{subfigure}
    \caption[]
    {\small Plot of the amplitude (dB) and phase (deg) of the $y$ component of the PWG electric field along the $y$-axis ($x=0$, $z=0.95\si{\meter}$) for multiple values of $\alpha=P_{inc/P_{S}}$. The two black vertical lines at $\pm R=\pm 0.24$ \si{\meter} delimit the QZ.} 
    \label{fig: PWG 1D}
\end{figure*}
As a consequence the constraint would be inactive as it is already satisfied by the unconstrained excitation vector $\bm{a}_U$.

\subsection{Plane Wave Generator: simulation results}
\label{sec: PWG simulation}

\begin{figure*}[t!]
  	\centering
   	\begin{subfigure}[b]{\columnwidth}
  	   	\centering
        \includegraphics{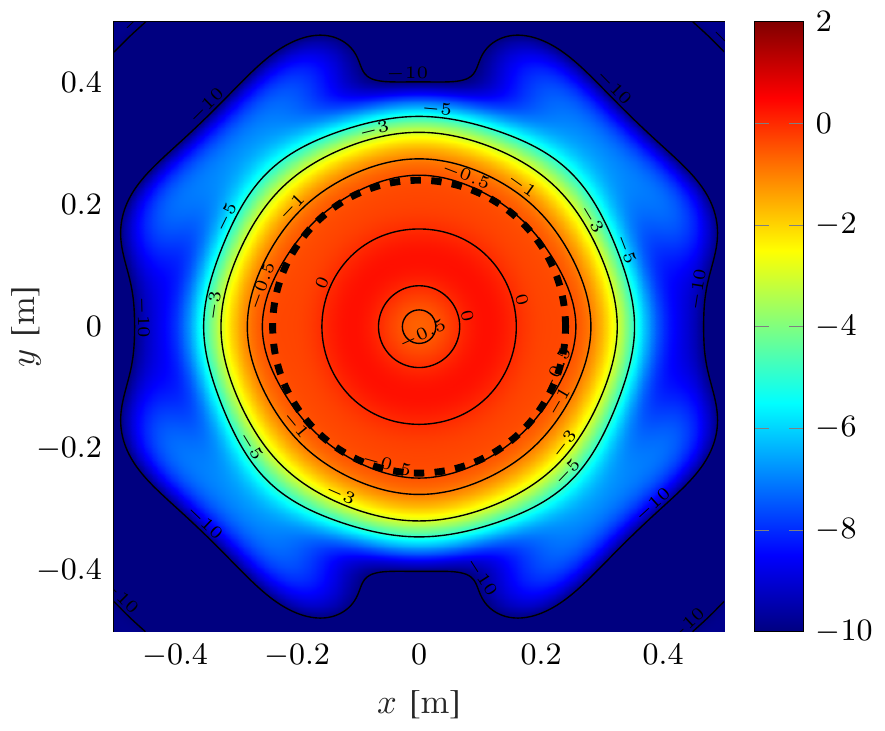}
        \caption[]%
        {{\small Amplitude ($x$-$y$ plane)}}    
        \label{fig: PWG 2D amp xy}
   	\end{subfigure}
    \hfill
    \begin{subfigure}[b]{\columnwidth}  
  	 	\centering 
        \includegraphics{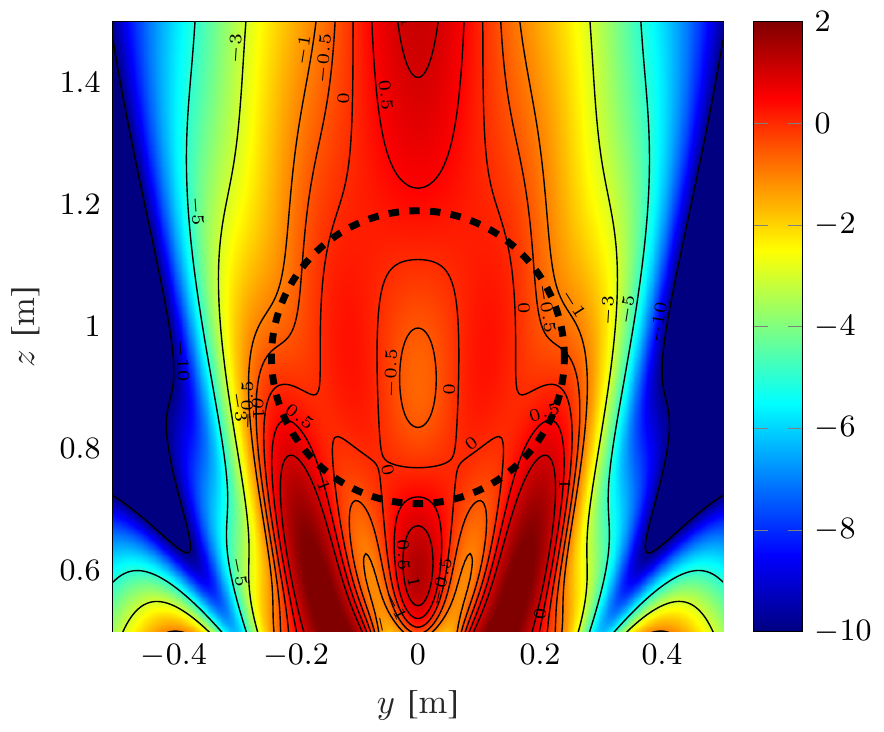}
        \caption[]%
        {{\small Amplitude ($y$-$z$ plane)}}    
        \label{fig: PWG 2D amp yz}
   	\end{subfigure}
    \vskip\baselineskip
    \begin{subfigure}[b]{\columnwidth}   
      	\centering 
        \includegraphics{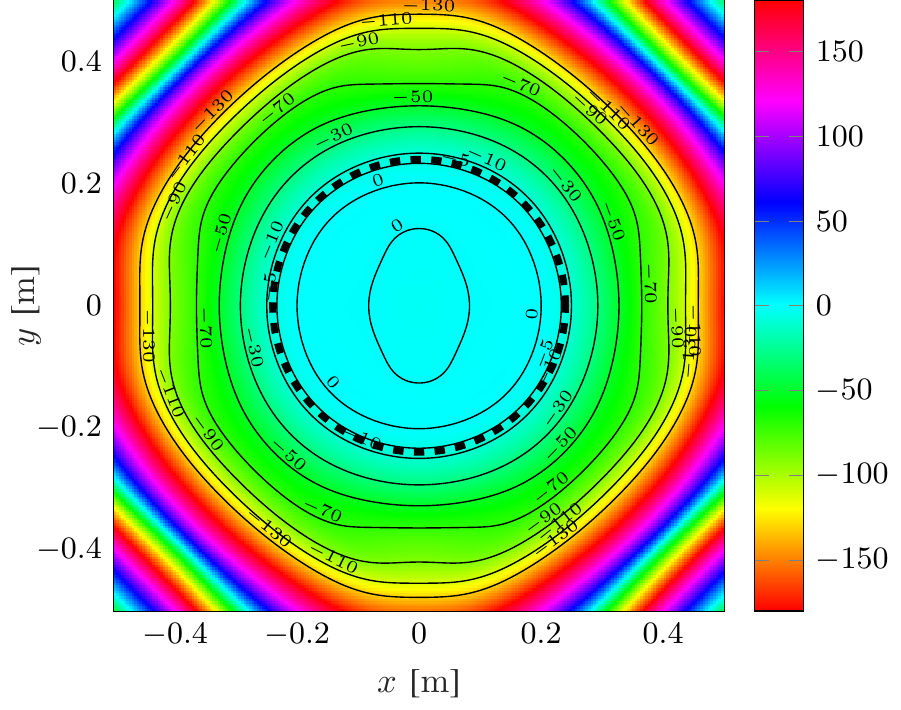}
        \caption[]%
        {{\small Phase ($x$-$y$ plane)}}    
        \label{fig: PWG 2D phase xy}
   	\end{subfigure}
    \hfill
    \begin{subfigure}[b]{\columnwidth}   
      	\centering 
        \includegraphics{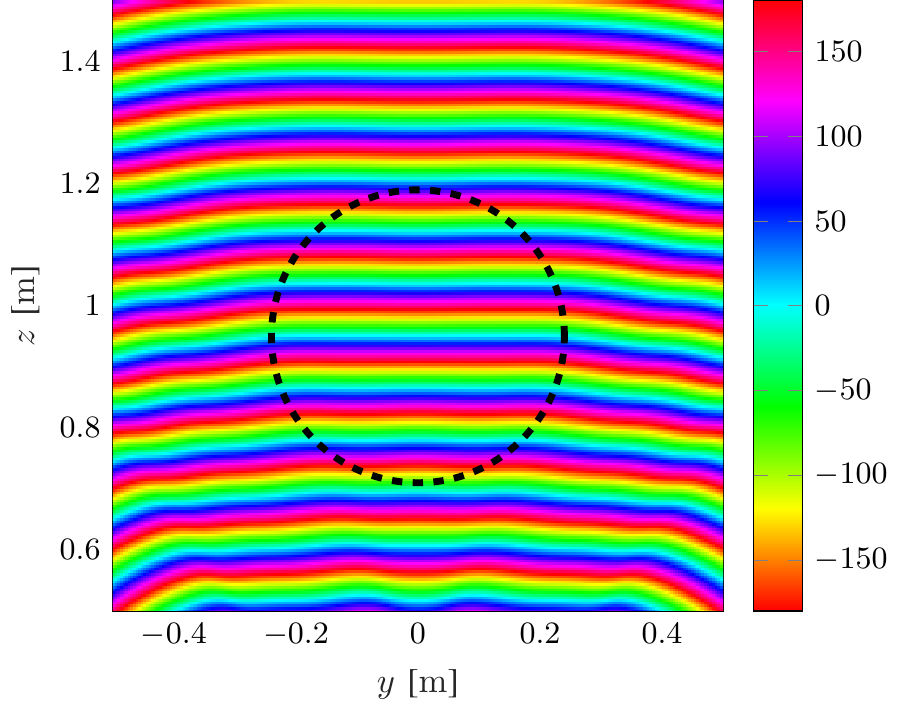}
        \caption[]%
        {{\small Phase ($y$-$z$ plane)}}    
        \label{fig: PWG 2D phase yz}
  	\end{subfigure}
    \caption[]
    {\small Plot of the amplitude (dB) and phase (deg) of the $y$ component of the PWG electric field in the $x$-$y$ and $y$-$z$ planes passing through the QZ center at $(0,0,0.95 \si{\meter})$ for $\alpha=P_{inc}/P_{S}=8.1$ dB. The plots in the $x$-$z$ plane are omitted since they do not significantly differ from the ones in the $y$-$z$ plane. The black dashed circle delimits the QZ.} 
    \label{fig: PWG 2D}
\end{figure*}

In this section we present the results obtained by simulating an array of $40$ elements arranged uniformly in $4$ concentrical rings as the one described in \cite{Scattone2019_1}. Let $n$ denote the index of each ring starting from the center. Then the number of elements per ring is equal to $4n$, and the ring radius is $R_n=nR_1$ with $R_1=80$ \si{\milli\meter}, $n=1,\ldots,4$. The array element is a $y$ oriented half-wave dipole placed $\lambda_0/4$ above an infinite ground plane resonating at $3.5$ \si{\giga\hertz} with length $39.5$ \si{\milli\meter}, radius $10^{-3}\lambda_0$ and excitation gap $10^{-3}\lambda_0$. The target region is a sphere of radius $R=240$ \si{\milli\meter} centered at $(0,0,z_0=950 \si{\milli\meter})$. As discussed in Sec. \ref{sec: PWG}, once the geometry of the system has been assigned the $\alpha$ value belongs to the interval $[\alpha_{min},\alpha_{max}]$. For this configuration $\alpha_{min}$ is equal to $3.0$, and $\alpha_{max}$ to $30.8$. To evaluate the performance of the PWG, we consider the following metric
\begin{equation}
	\label{eq: PWG relative error}
	\begin{split}
	\epsilon^2\left(\bm{E},\bar{\bm{E}}\right) &\triangleq \frac{\norm{(\Delta\bm{E},\Delta\bm{H})}^2}{\norm{(\bm{\bar{E}},\bm{\bar{H}})}^2}=\frac{\iiint_{V} \frac{1}{4}\norm{\Delta\bm{E}}^2 d\tau}{\iiint_{V} \frac{1}{4} \norm{\bar{\bm{E}}}^2 d\tau}=\\
	&=\frac{\iiint_{V} \norm{\bm{E}-\bar{\bm{E}}}^2 d\tau}{\iiint_{V} \norm{\bar{\bm{E}}}^2 d\tau},		
	\end{split}
\end{equation}
that we call relative error and represents the ratio between the electric energy stored in the volume $V$ by the error field $\bm{\Delta E}$ and by the target field $\bar{\bm{E}}$. This metric is significant for any target field, not only uniform ones, and takes into account both phase and amplitude of the field. Two other common metrics used to evaluate the performance of a PWG are the maximum amplitude and phase deviation of the electric field. Fig. \ref{fig: PWG norm} shows the relative error as a function of $\alpha$, with both quantities expressed in dB. Since $\alpha$ represents the ratio between the incident power $P_{inc}$ and the active power flow through the surface $S$ $P_S$, the higher the $\alpha$ the lower the efficiency. It is evident that the relative error decreases monotonically when $\alpha$ increases. To better understand this behaviour, we consider Fig. \ref{fig: PWG 1D}, which shows the amplitude and phase of the $y$ component of the electric field along the y axis when $x=0$ and $z=z_0=950$ \si{\milli\meter} for four values of $\alpha$. When $\alpha$ is equal to $5.7$ dB the field is more concentrated inside the QZ, but both the amplitude and phase deviation are higher compared to the other cases. When $\alpha$ grows the field decreases less sharply outside the QZ, and both amplitude and phase ripples decrease. When $\alpha$ is $13$ dB, on one hand the peak of the field is outside the QZ, meaning that most of the power flows outside the QZ, on the other hand the ripple amplitude is the lowest. Fig. \ref{fig: PWG 2D} shows the amplitude and phase of the $y$ component of the electric field in the $x$-$y$ and $y$-$z$ planes cutting the QZ at its center for $\alpha=8.1$ dB. The results for the $x$-$z$ plane are similar to the ones in the $y$-$z$ plane and have been omitted for brevity. Considering the three cutting planes passing through the QZ center, the maximum amplitude deviation is $0.84$ dB, and the maximum phase deviation is $7.1$ deg. These results are in good agreement with the ones presented in \cite{Scattone2019_2}. Although the two array geometries are similar, in \cite{Scattone2019_2} a wideband antenna is used as element, while in the present work a half-wave dipole is used.\\
\indent Finally it is worth noting that Fig. \ref{fig: PWG norm} could be used for the design of the PWG excitations, as a trade-off between the PWG performance in the assigned QZ and the percentage of power that is actually flowing through the target region. Similar curves can be obtained by considering the maximum amplitude and phase deviation as a function of $\alpha$. Once the specifications for the field ripple in the QZ have been assigned, one can easily find the minimum $\alpha$ value that allows to meet them.


\section{Conclusions}
\label{sec: Conclusions} 

The concept of inner product has been applied to electromagnetic fields to synthesize the excitations of near-field arrays. From the linearity of the problem, the EM field generated by the array can be expanded into a basis obtained by feeding the array with a set of orthogonal excitations. By exploiting the properties of inner products and their induced norm, two different methods to compute the excitations have been introduced and discussed. The "maximum norm" method provides the excitation vector that maximizes the induced norm for a given input power. In Sec. \ref{sec: maximum active power flow through a surface} a proper selection of the EM inner product led to the maximum power flow method. The latter has been compared with the conjugate phase method, and the two give very close performance when the target surface is electrically small, as expected. Since the maximum power flow method can be applied to a surface located in any region of the antenna surrounding, it has been verified that it converges to the "maximum BCE" method when applied to a target surface in the far-field region of the array. The "minimum error field norm" method can be applied for the synthesis of a target field. In Sec. \ref{sec: PWG}, as an example, the constrained variant of the method has been applied to optimize the excitations of a Plane Wave Generator.\\ 
\indent On the one hand, both methods provide a closed-form solution for the array excitations; on the other hand, to obtain the matrices and vectors needed to compute the solution, one has to numerically solve surface or volume integrals, depending on the chosen inner product, involving the EM fields of the basis. If the target region is in the far-field of the array, and the effects of mutual coupling can be neglected, then analytical expressions for the radiated fields can be used, as described in Sec. \ref{sec: maximum active power flow: Far-Field case}. Otherwise, the EM field must be obtained via simulations or measurements which may be quite time-consuming. Finally, the synthesis methods here discussed are based on global properties of the EM fields, such as the maximization of the active power flow through a surface, rather than on specific local conditions assigned at a grid of points in the antenna near-field region.

\appendices


\section{EM field of a linear system}
\label{ap: EM field of a linear system}

If we consider an N-port antenna system, we can explicitly express the dependence of the EM field on the excitation vector $\bm{a} \in \mathbb{C}^{N}$ using the notation $\bm{E}(\bm{a},\bm{r})$, $\bm{H}(\bm{a},\bm{r})$. If we restrict the discussion to linear systems, then we can express the electric field generated by the excitation vector $\bm{a}= \sum_{n=1}^N \zeta_n \bm{x}_n$ as
\begin{equation}
	\label{def: linear EM system}
	\bm{E}(\bm{a},\bm{r})=\bm{E}\left(\sum_{n=1}^N \zeta_n \bm{x}_n,\bm{r}\right)=\sum_{n=1}^N \zeta_n \bm{E}\left(\bm{x}_n,\bm{r}\right),
\end{equation}
where $\{\zeta_n \in \mathbb{C}\}_{n=1}^N$ is a set of scalar complex values, and $\{\bm{x}_n \in \mathbb{C}^{N}\}_{n=1}^N$ is a set of linearly independent vectors in $\mathbb{C}^{N}$. Since any set of $N$ linearly independent vectors forms a basis in $\mathbb{C}^{N}$, then we can express any vector in the space as a linear combination of the basis 
\begin{equation}
	\bm{a}=\sum_{n=1}^N \zeta_n \bm{x}_n=\bm{X}\bm{\zeta},
\end{equation}
with $\bm{X}\triangleq [\bm{x}_1, \cdots ,\bm{x}_N] \in \mathbb{C}^{N \times N}$ and $\bm{\zeta}\triangleq [\zeta_1, \cdots ,\zeta_N]^T\in \mathbb{C}^{N}$. From the linear independence of its column vectors, it follows that $\bm{X}$ is invertible, so 
\begin{equation}
	\label{eq: a from zeta}
	\bm{\zeta}=\bm{X}^{-1}\bm{a}.
\end{equation}
Substituting (\ref{eq: a from zeta}) into (\ref{def: linear EM system}) and defining $\bm{\mathcal{E}}_n(\bm{r})\triangleq \bm{E}(\bm{x}_n,\bm{r})$ for $n=1,\dots,N$, we obtain
\begin{equation}
	\label{eq: EM field of a expanded in a orthogonal basis}
	\bm{E}(\bm{a},\bm{r})=\sum_{n=1}^N \left( [\bm{X}^{-1}]_n \bm{a} \right) \bm{\mathcal{E}}_n(\bm{r}).
\end{equation}
If we consider an orthonormal basis, then $\bm{X}$ is a unitary matrix, and (\ref{eq: EM field of a expanded in a orthogonal basis}) becomes
\begin{equation}
	\label{eq: electric field of a expanded in a orthonormal basis}
	\bm{E}(\bm{a},\bm{r})=\sum_{n=1}^N \left( \bm{\hat{x}}_n^H \bm{a} \right) \bm{\mathcal{E}}_n(\bm{r}).
\end{equation}

By following the same steps with the magnetic field, we obtain
\begin{equation}
	\label{eq: magnetic field of a expanded in a orthonormal basis}
	\bm{H}(\bm{a},\bm{r})=\sum_{n=1}^N \left( \bm{\hat{x}}_n^H \bm{a} \right) \bm{\mathcal{H}}_n(\bm{r}),
\end{equation}
with $\bm{\mathcal{H}}_n(\bm{r})\triangleq \bm{H}(\bm{\hat{x}}_n,\bm{r})$, $n=1,\dots,N$.


\section{Solution of the optimization problem in Sec. \ref{sec: Maximum norm method}}
\label{ap: Solution of the constrained optimization problem }

In this appendix we derive the solution of the constrained optimization problem in Sec. \ref{sec: Maximum norm method}.
\begin{IEEEproof}
To solve the constrained optimization problem we exploit the method of Lagrange multipliers. The Lagrangian function can be expressed as
\begin{equation}
	\label{eq: lagrangian function complex}
	\begin{split}
		\mathcal{L}(\bm{a},\mu)=& \bm{a}^H\bm{A}\bm{a} - \mu( \bm{a}^H\bm{B}\bm{a} - \bar{P} )=\\
		=&\bm{a}^H ( \bm{A} - \mu\bm{B} ) \bm{a} + \mu\bar{P},
	\end{split}
\end{equation}
which is a real function of the complex variable $\bm{a}$. If we define $\Re\{\bm{a}\}\triangleq \bm{a}_r$ and $\Im\{\bm{a}\}\triangleq \bm{a}_i$, the Lagrangian function becomes
\begin{equation}
	\label{eq: lagrangian function real}
	\mathcal{L}(\bm{a}_r,\bm{a}_i,\mu)=(\bm{a}_r^T-j\bm{a}_i^T) ( \bm{A} - \mu\bm{B} ) (\bm{a}_r+j\bm{a}_i) + \mu\bar{P}.
\end{equation}
By computing the partial derivative with respect to each element of $\bm{a}_r$ and $\bm{a}_i$, we get
\begin{equation}
	\label{eq: stationary point scalar}
	\begin{split}
		&\frac{\partial}{\partial [\bm{a}_r]_n} \mathcal{L}(\bm{a}_r,\bm{a}_i,\mu)=2\Re\{\hat{e}^T_n ( \bm{A} - \mu\bm{B} ) (\bm{a}_r+j\bm{a}_i)\}=0,\\
		&\frac{\partial}{\partial [\bm{a}_i]_n} \mathcal{L}(\bm{a}_r,\bm{a}_i,\mu)=2\Im\{\hat{e}^T_n ( \bm{A} - \mu\bm{B} ) (\bm{a}_r+j\bm{a}_i)\}=0,
	\end{split}
\end{equation}
for $n=1,\dots,N$, which can be expressed in the following more compact form
\begin{equation}
	\label{eq: stationary point vector}
	( \bm{A} - \mu\bm{B} ) \bm{a}=\bm{0}.
\end{equation}
(\ref{eq: stationary point vector}) corresponds to a generalized eigenvalue problem, and $\{(\bm{\hat\vartheta}_n,\mu_n)\}_{n=1}^N$ are the associated eigenvectors and eigenvalues. By left-multiplying (\ref{eq: stationary point vector}) by $\bm{a}^H$ and considering the constraint (\ref{eq:constraint1}), it follows that
\begin{equation}
	\bm{a}^H\bm{A}\bm{a}=\mu(\bm{a}^H\bm{B}\bm{a})=\mu \bar{P},
\end{equation}
thus the excitation vector that maximizes (\ref{eq:optProb}) must be \mbox{$\bm{a}_{opt}=c\bm{\hat\vartheta}_{max}, c \in \mathbb{C}$}, where $\bm{\hat\vartheta}_{max}$ is the eigenvector associated to the highest eigenvalue $\mu_{max}$. By imposing the constraint (\ref{eq:constraint1}), we find that $c$ must satisfy the following equation
\begin{equation}
	\vert c \vert=\sqrt{\frac{\bar{P}}{\bm{\hat\vartheta}^H_{max}\bm{B}\bm{\hat\vartheta}_{max}}},
\end{equation}
so (\ref{eq: a opt norm}) is a suitable solution.
\end{IEEEproof}


\section{Solution of the optimization problem in Sec. \ref{sec: Minimum error field norm method} }
\label{ap: Solution of the minimum error norm optimization problem}

In this appendix we derive the solution of the optimization problem in Sec. \ref{sec: Minimum error field norm method}
\begin{IEEEproof}
Let us start by explicating the norm squared of the error field as follows
\begin{equation}
	\label{eq: error field norm (a explicit)}
	\begin{split}
	\norm{(\Delta \bm{E},\Delta \bm{H})}^2=&\norm{(\bm{\bar{E}},\bm{\bar{H}}) - \sum_{n=1}^N \left( \bm{\hat{x}}_n^H \bm{a} \right) (\bm{\mathcal{E}}_n,\bm{\mathcal{H}}_n)}^2=\\
	=&\norm{(\bm{\bar{E}},\bm{\bar{H}})}^2+\norm{\sum_{n=1}^N \left( \bm{\hat{x}}_n^H \bm{a} \right) (\bm{\mathcal{E}}_n,\bm{\mathcal{H}}_n)}^2 - \\
	&- 2\Re\left\{ \inner{(\bm{\bar{E}},\bm{\bar{H}})}{\sum_{n=1}^N \left( \bm{\hat{x}}_n^H \bm{a} \right) (\bm{\mathcal{E}}_n,\bm{\mathcal{H}}_n)} \right\},
	\end{split}
\end{equation}
where we exploited the following equality
\begin{equation}
	\label{eq: norm squared of the difference}
	\begin{split}
	\norm{u_1-u_2}^2=&\inner{u_1-u_2}{u_1-u_2}=\\
	=&\norm{u_1}^2+	\norm{u_2}^2-\inner{u_1}{u_2}-\inner{u_2}{u_1}=\\
	=&\norm{u_1}^2+	\norm{u_2}^2-2\Re \left\{ \inner{u_1}{u_2} \right\},
	\end{split}
\end{equation}
that holds for any pair $(u_1,u_2)$ belonging to an inner product space. We now proceed by expressing the second and third terms in the last equality in (\ref{eq: error field norm (a explicit)}) in vector form. The second term is exactly equal to the one in (\ref{eq: norm}), while the third term corresponds to
\begin{equation}
	\begin{split}
	\label{eq: third term vector}
		\inner{(\bm{\bar{E}},\bm{\bar{H}})}{&\sum_{n=1}^N \left( \bm{\hat{x}}_n^H \bm{a} \right) (\bm{\mathcal{E}}_n,\bm{\mathcal{H}}_n)}=\\
		=& \sum_{n=1}^N \left( \bm{a}^H \bm{\hat{x}}_n \right) \inner{(\bm{\bar{E}},\bm{\bar{H}})}{(\bm{\mathcal{E}}_n,\bm{\mathcal{H}}_n)}=\\
		=& \bm{a}^H \bm{X} \bm{v}.
	\end{split}
\end{equation}
Substituting (\ref{eq: norm}) and (\ref{eq: third term vector}) into (\ref{eq: error field norm (a explicit)}) yields
\begin{equation}
	\label{eq: error field norm (a explicit) final}
		\norm{(\Delta \bm{E},\Delta \bm{H})}^2=\norm{(\bm{\bar{E}},\bm{\bar{H}})}^2 + \bm{a}^H \bm{A} \bm{a} - 2\Re \left\{ \bm{a}^H \bm{X} \bm{v} \right\}.
\end{equation}
The solution of the constrained optimization problem has to satisfy the following conditions (Karush-Kuhn-Tucker conditions):

\begin{enumerate}
  \item stationarity of \mbox{$\mathcal{L}(\bm{a},\xi)$},
  \item $\bm{a}^H\bm{C}\bm{a}-h \leq 0$,
  \item $\xi \geq 0$,
  \item $\xi (\bm{a}^H\bm{C}\bm{a}-h) = 0$,
\end{enumerate}
where we have defined the following function
\begin{equation}
\begin{split}
	\mathcal{L}(\bm{a},\xi) \triangleq & \norm{(\Delta \bm{E},\Delta \bm{H})}^2 + \xi (\bm{a}^H\bm{C}\bm{a}-h) =\\
	=& \bm{a}^H (\bm{A} + \xi \bm{C} )\bm{a} - 2\Re \left\{ \bm{a}^H \bm{X} \bm{v} \right\}- \xi h +\norm{(\bm{\bar{E}},\bm{\bar{H}})}^2.
\end{split}
\end{equation}
In order to find the stationary points of $\mathcal{L}(\bm{a},\xi)$, we have to compute its partial derivatives with respect to the real and imaginary part of each element of $\bm{a}$. The procedure is almost identical to the one found in Appendix \ref{ap: Solution of the constrained optimization problem } and leads to the following vectorial equation
\begin{equation}
	(\bm{A} + \xi \bm{C} )\bm{a}=\bm{X} \bm{v},
\end{equation}
so Condition (1) is equivalent to the condition

\begin{enumerate}[label={\arabic*')}]
\Item 
	\begin{align}
	\label{eq: a opt constrained mimimum error norm}
		\bm{a}=(\bm{A}+\xi\bm{C})^{-1}\bm{X}\bm{v},
	\end{align}
\end{enumerate}
if  $\bm{A} + \xi \bm{C}$ is invertible.
In order to satisfy Condition (4) either $\xi$ or $\bm{a}^H\bm{C}\bm{a}-h$ must be equal to $0$. We will analyse the case $\xi=0$ first, and the  case $\xi > 0$ later.

\noindent\textbf{Case 1: $\bm{\xi=0$}}

Since $\xi=0$ Condition (3) and (4) are satisfied, and Condition (1') becomes 
\begin{equation}
	\bm{a}_U=\bm{A}^{-1}\bm{X} \bm{v},
\end{equation}
which is exactly the solution of the unconstrained optimization problem \ref{eq: minimum error norm a general basis}.
Substituting the latter expression in Condition (2) leads to the following inequality
\begin{equation}
	\bm{a}_U^H\bm{C}\bm{a}_U-h = (\bm{A}^{-1}\bm{X} \bm{v})^H\bm{C}(\bm{A}^{-1}\bm{X} \bm{v})-h  \leq 0.
\end{equation}
This case corresponds to the one where the solution of the unconstrained optimization problem already satisfies the constraint.

\noindent\textbf{Case 2: $\bm{\xi>0$}}

Since $\xi>0$ Condition (3) is satisfied. Condition (2) and (4) are satisfied if and only if $\bm{a}^H\bm{C}\bm{a}-h=0$. By substituting Condition (1') in the latter equation we obtain the following non-linear equation for $\xi$
\begin{equation}
	\label{eq: xi equation}
	\begin{split}
		0 \labelrel={rel: xi a} &\bm{v}^H\bm{X}^H(\bm{A} + \xi \bm{C} )^{-1} \bm{C} (\bm{A} + \xi \bm{C} )^{-1}\bm{X} \bm{v} -h= \\
		\labelrel={rel: xi b} &\bm{v}^H\bm{X}^H\bm{G}^{-1}(\bm{G}^{-1}\bm{A}\bm{G}^{-1} + \xi \bm{I}_N )^{-2} \bm{G}^{-1}\bm{X} \bm{v} -h=\\
		\labelrel={rel: xi c} &\bm{v}^H\bm{X}^H\bm{G}^{-1}\bm{V}(\bm{D} + \xi \bm{I}_N )^{-2} \bm{V}^{-1}\bm{G}^{-1}\bm{X} \bm{v} -h=\\
		\labelrel={rel: xi d} & \bm{w}_1^H(\bm{D} + \xi \bm{I}_N )^{-2}\bm{w}_2-h=\sum_{n=1}^N \frac{[\bm{w}_1]_n^*[\bm{w}_2]_n}{([\bm{D}]_{n,n}+\xi)^2}-h,
	\end{split}
\end{equation}
where we have used the fact that $\bm{A} + \xi \bm{C}$ is hermitian in ~\eqref{rel: xi a}. Since $\bm{C}$ is hermitian its eigenvalue decomposition corresponds to $\bm{V'}\bm{D'}\bm{V'}^H$, where $\bm{V'}$ is a unitary matrix, and $\bm{D'}$ is a real diagonal matrix. In ~\eqref{rel: xi b} we have defined $\bm{G}\triangleq \bm{V'}\bm{D'}^{\frac{1}{2}}\bm{V'}^H$ so $\bm{C}=\bm{G}\bm{G}$, where $\bm{D'}^{\frac{1}{2}}$ is a diagonal matrix with $[\bm{D'}^{\frac{1}{2}}]_{n,n}=\sqrt{[\bm{D'}]_{n,n}}$. Since the square root admits two solutions, the matrix $\bm{G}$ is not uniquely defined, but either choice leads to the same solution, except for numerical errors. In ~\eqref{rel: xi c} we have used the eigenvalue decomposition $\bm{G}^{-1}\bm{A}\bm{G}^{-1}=\bm{V}\bm{D}\bm{V}^{-1}$. Finally in ~\eqref{rel: xi d} we have defined the two vectors $\bm{w}_1 \triangleq \bm{V}^H(\bm{G}^{-1})^H\bm{X}\bm{v}$ and $\bm{w}_2 \triangleq \bm{V}^{-1}\bm{G}^{-1}\bm{X}\bm{v}$. Eq. \ref{eq: xi equation} can be recast into a polynomial equation of order $2N$ by multiplying both terms by $\prod_{n=1}^N ([\bm{D}]_{n,n}+\xi)^2$. As a consequence the set of solutions $\Psi_{\xi}$ contains $2N$ complex values. Let $\Psi_{\xi}^+ \subseteq \Psi_{\xi}$ be the subset containing the real positive solutions. Since $\xi$ must satisfy Condition (3), the optimal value corresponds to the one that minimizes $\norm{(\Delta \bm{E},\Delta \bm{H})}^2$ among the ones in $\Psi_{\xi}^+$. Once the optimal $\xi$ value is found the associated excitation vector can be computed using (\ref{eq: a opt constrained mimimum error norm}).

\end{IEEEproof}


\section{Proof of the properties of the EM inner product}
\label{ap: Proof of the properties of the EM inner product}

In this appendix we prove that the operator in \mbox{Definition \ref{def: EM inner product}} satisfies all the properties of an inner product. Given $(\bm{E}_{1},\bm{H}_{1}),(\bm{E}_{2},\bm{H}_{2}),(\bm{E},\bm{H}) \in \mathcal{O}(S)$ and $\lambda_1,\lambda_2 \in \mathbb{C}$, the three following properties must be satisfied \cite{Lang1987}:

\begin{enumerate}
	\item \textit{conjugate symmetry}: 
	\begin{equation*}
		\inner{(\bm{E}_{1},\bm{H}_{1})}{(\bm{E}_{2},\bm{H}_{2})}=\inner{(\bm{E}_{2},\bm{H}_{2})}{(\bm{E}_{1},\bm{H}_{1})}^*.		
	\end{equation*} 
	\item \textit{Linearity in the first argument}:
	\begin{equation*}
		\begin{split}
			&\inner{\lambda_1( \bm{E}_{1},\bm{H}_{1})+\lambda_2( \bm{E}_{2},\bm{H}_{2})}{(\bm{E},\bm{H})}=\\
			&=\lambda_1\inner{( \bm{E}_{1},\bm{H}_{1})}{(\bm{E},\bm{H})}+\lambda_2\inner{( \bm{E}_{2},\bm{H}_{2})}{(\bm{E},\bm{H})}.
		\end{split}
	\end{equation*} 
	\item \textit{Positive-definiteness}:
	\begin{equation*}
		\inner{( \bm{E},\bm{H})}{(\bm{E},\bm{H})} \triangleq \norm{(\bm{E},\bm{H})}^2 \geq 0,
	\end{equation*} 
	where the equality $\norm{(\bm{E},\bm{H})}^2 = 0$ holds if and only if $(\bm{E},\bm{H})=(\bm{0},\bm{0}), \forall \bm{r} \in S$.
\end{enumerate}

\begin{IEEEproof}
	Property 1 and 2 are a direct consequence of the presence of the conjugate operator in the second argument and the linearity of the integral operator, respectively. While property 3 follows from the fact that
	\begin{equation*}
		\begin{split}
			\norm{(\bm{E},\bm{H})}^2 &\triangleq \inner{( \bm{E},\bm{H})}{(\bm{E},\bm{H})}\\
			&=\frac{1}{2} \iint_{S} \Re \big\{ \bm{E} \times \bm{H}^* \big\} \cdot \bm{\hat{n}}d\Sigma \geq 0,
		\end{split}
	\end{equation*}
	where the last inequality holds by Definition \ref{def: O set}.
\end{IEEEproof}


\ifCLASSOPTIONcaptionsoff
  \newpage
\fi

\bibliographystyle{IEEEtran}
\bibliography{main}

\begin{IEEEbiography}[{\includegraphics[width=1in,height=1.25in,clip,keepaspectratio]{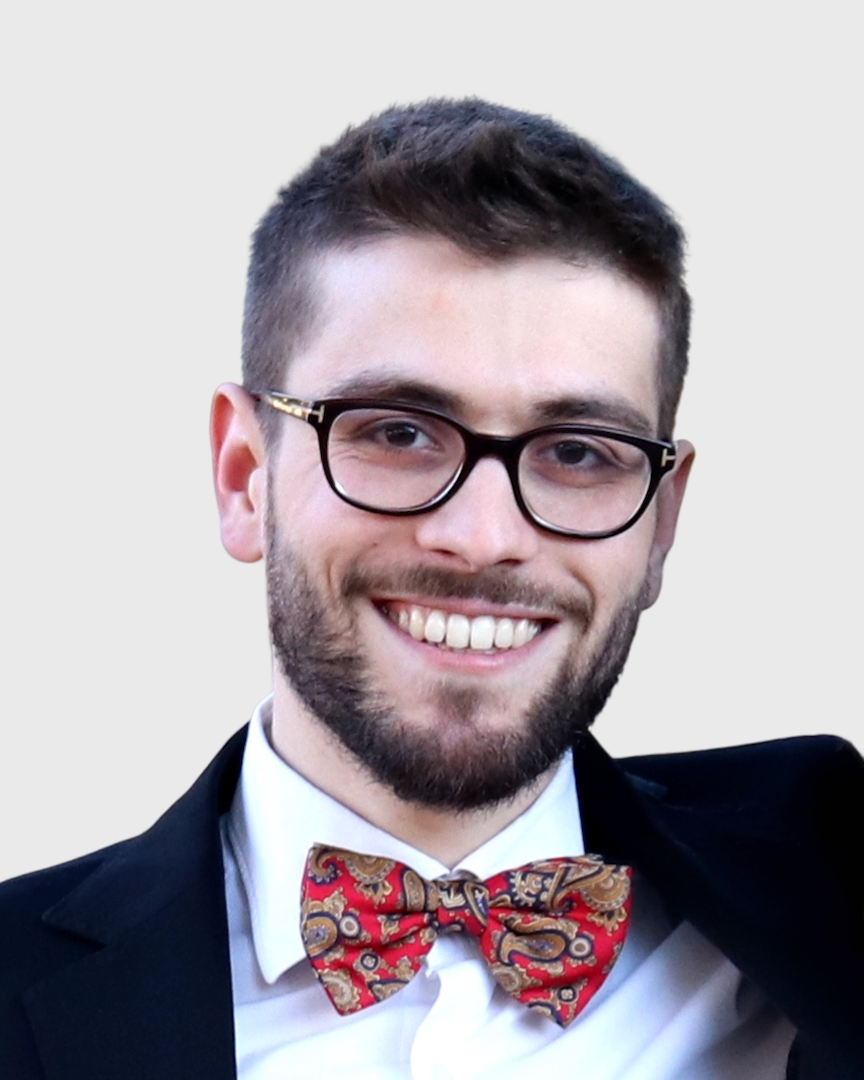}}]{Francesco Lisi}
(Student Member, IEEE) received the Bachelor Degree (cum laude) in electronics and telecommunications engineering from University of Florence in 2019 and the Master Degree (cum laude) in telecommunications engineering from University of Pisa in 2021. He is currently pursuing the PhD degree in information engineering with the University of Pisa. In April-September 2021 he was a student intern at the Université Paris-Saclay, CNRS, CentraleSupélec, Laboratoire des Signaux et systèmes (L2S), where he worked on his master thesis project on the development of a reinforcement learning based algorithm for massive MIMO radar systems. He was the recipient of the 'Renato Mariani' prize awarded by the Italian Association of Electrical Engineering, Electronics, Automation, Informatics and Telecommunications (AEIT) in 2022. \\
His research interests include antenna arrays, near-field focusing, wireless power transfer systems, MIMO radars and reinforcement learning algorithms.
\end{IEEEbiography}

\begin{IEEEbiography}[{\includegraphics[width=1in,height=1.25in,clip,keepaspectratio]{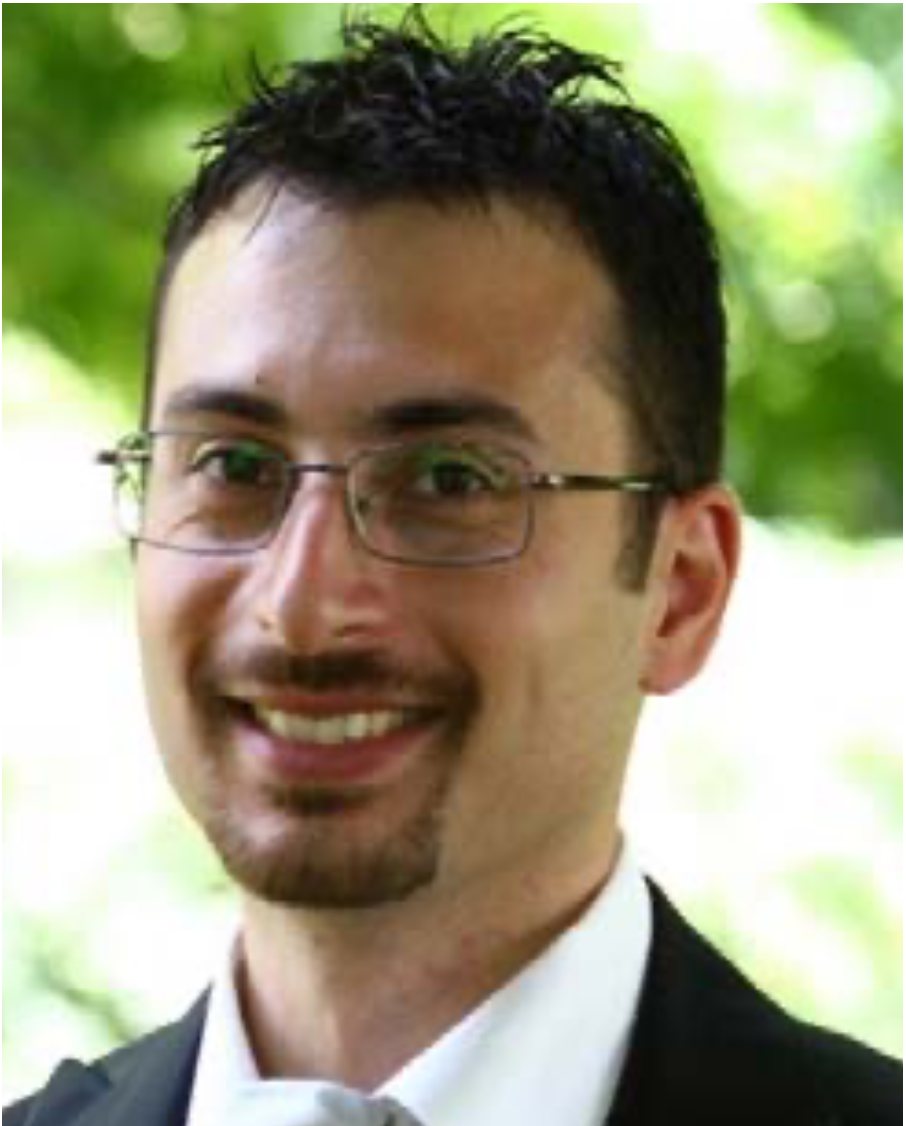}}]{Andrea Michel}
(Senior Member, IEEE) received the B.E., M.E. and Ph.D. degrees in Telecommunications Engineering from the University of Pisa, Italy, in 2009, 2011, and 2015, respectively. In 2014, he was a Visiting Scholar with the Electro Science Laboratory, The Ohio State University, Columbus, OH, USA, under the supervision of Prof John Volakis. During this period, he was involved in research on a theoretical analysis on the accuracy of a novel technique for deep tissue imaging. Since 2015, he has been a Post-Doctoral Researcher in Applied Electromagnetism at the Microwave and Radiation Laboratory, Department of Information Engineering, University of Pisa, where he is currently an Assistant Professor. He is involved in the design of antennas for automotive applications, MIMO systems, and wearable communication systems, also in collaboration with other research institutes and companies. His current research interests include the design of integrated antenna for communication systems and smart antennas for near field UHF-RFID readers. Dr. Michel was a recipient of the Young Scientist Award from the International Union of Radio Science (URSI), Commission B, in 2014, 2015, and 2016. In 2016, he received the Best Paper Honorary Mention from the IEEE International Conference on RFID Technology and Applications, Shunde, Guangdong, China. He is Early Career Representative for URSI Commission B (Fields and Waves). He serves as Associate Editor for URSI Radio Science Letters and URSI Radio Science Bulletin journals.
\end{IEEEbiography}


\begin{IEEEbiography}[{\includegraphics[width=1in,height=1.25in,clip,keepaspectratio]{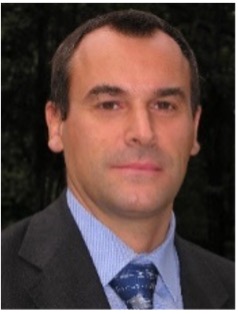}}]{Paolo Nepa}
(Senior Member, IEEE) received the Laurea Degree in electronics engineering (summa cum laude) from the University of Pisa, Italy, in 1990. Since 1990, he has been with the Department of Information Engineering, University of Pisa, where he is currently a Full Professor. In 1998, he was at the Electro Science Laboratory (ESL), The Ohio State University (OSU), Columbus, OH, as a Visiting Scholar supported by a grant of the Italian National Research Council. At the ESL, he was involved in research on efficient hybrid techniques for the analysis of large antenna arrays. His research interests include the extension of high-frequency techniques to electromagnetic scattering from material structures and its application to the development of radio propagation models for indoor and outdoor scenarios of wireless communication systems. He is also involved in the design of wideband and multiband antennas for mobile communication systems, as well as in the design of antennas optimized for near-field coupling and focusing. He was working on channel characterization, wearable antenna design and diversity scheme implementation, for body-centric communication systems. In the context of UHF-RFID systems, he is working on efficient techniques for radiolocalization of either tagged objects or mobile readers in the context of IoT and Smart Industry scenarios. He has co-authored more than 300 international journal articles and conference contributions.\\
Since 2013 he is a member of the Technical Advisory Board of URSI Commission B - Fields and Waves. He served as TPC member of several international IEEE conferences. In 2019, he has been the General Chair of the IEEE RFID-TA 2019 International Conference. Since 2016, he has been serving as an Associate Editor for the IEEE Antennas and Wireless Propagation Letters and in 2021 he was a recipient of the Outstanding Associate Editors Awards. Since 2021, he is an Associate Editor for the IEEE Transactions on Antennas and Propagation. He was a recipient of the Young Scientist Award from the International Union of Radio Science, Commission B, in 1998.
\end{IEEEbiography}




\end{document}